\documentclass[12pt,preprint]{aastex}

\slugcomment{accepted for publication in ApJ on Aug. 11, 2003}

\shorttitle{Dust Formation in Pop. III Supernovae}
\shortauthors{Nozawa et al.}

\begin{document}


\title{Dust in the Early Universe : Dust Formation \\ in the Ejecta of 
Population III Supernovae }


\author{Takaya Nozawa and Takashi Kozasa}
\affil{Division of Earth and Planetary Sciences, Graduate School of 
Science, Hokkaido University, Sapporo 060-0810, Japan; 
nozawabozo@ep.sci.hokudai.ac.jp, kozasa@ep.sci.hokudai.ac.jp}

\and

\author{Hideyuki Umeda, Keiichi Maeda, and Ken'ichi Nomoto}
\affil{Department of Astronomy, School of Science, 
University of Tokyo, Bunkyo-ku, Tokyo 113-0033, Japan; 
umeda@astron.s.u-tokyo.ac.jp, maeda@astron.s.u-tokyo.ac.jp, 
nomoto@astron.s.u-tokyo.ac.jp}



\begin{abstract}
Dust grains play a crucial role on formation and evolution history of
stars and galaxies in the early universe. 
We investigate the formation of dust grains in the ejecta of population 
III supernovae including pair--instability supernovae which are expected
to occur in the early universe, applying a theory of non--steady state 
nucleation and grain growth. 
Dust formation calculations are performed for core collapse supernovae 
with the progenitor mass $M_{\rm pr}$ ranging from 13 to 30 $M_{\odot}$ 
and for pair--instability supernovae with $M_{\rm pr}$ = 170  and 200 
$M_{\odot}$. 
In the calculations, the time evolution of gas temperature in the
ejecta, which strongly affects the number density and size of newly
formed grains, is calculated by solving the radiative transfer equation 
taking account of the energy deposition of radio active elements. 
Two extreme cases are considered for the elemental composition 
in the ejecta; unmixed and uniformly mixed cases within the
He--core, and formation of CO and SiO molecules is assumed to be 
complete.

The results of calculations for core collapse supernovae and
pair--instability supernovae are summarized as the followings; 
in the unmixed ejecta, a variety of grain species condense, reflecting 
the difference of the elemental composition at the formation site in the 
ejecta, otherwise only oxide grains condense in the uniformly mixed
ejecta. 
The average size of newly formed grains spans the range of three orders 
of magnitude, depending on the grain species and the formation
condition, and the maximum radius is limited to less than 1 $\mu$m,
which does not depend on the progenitor mass. 
The size distribution function of each grain species is approximately 
log--normal, except for Mg--silicates, MgO, Si and FeS grains in the 
unmixed case and Al$_2$O$_3$ grain in the both cases. 
The size distribution function summed up over all grain species is
approximated by a power--law formula whose index is $-3.5$ for the
larger radius and $-2.5$ for the smaller one; the radius at the
crossover point ranges from 0.004 to 0.1 $\mu$m, depending on the model 
of supernovae. 
The fraction of mass locked into dust grains increases with increasing 
the progenitor mass; 2--5 \% of the progenitor mass for core collapse 
supernovae and 15--30 \% for pair--instability supernovae whose
progenitor mass ranges from 140 to 260 $M_{\odot}$. 
Thus, if the very massive stars populate the first generation stars 
(population III stars), a large amount of dust grains would be produced in
the early universe. 
We also discuss the dependence of the explosion energy and the amount of
$^{56}$Ni in the ejecta as well as the efficiency of formation of CO and
SiO molecules on the formation of dust grains in the ejecta of supernovae.
\end{abstract}




\section{Introduction}

The recent observations of the reddening of background quasars and the 
damped Ly$\alpha$ systems in the spectra of distant quasars have
confirmed the presence of dust grains at high--redshifted universe 
(e.g., Pettini et al. 1994, 1997; Pei \& Fall 1995). 
Dust grains play a critical role on formation and evolution history of 
stars and galaxies in the early universe: 
Dust grains absorb stellar light and re--emit it by thermal radiation, 
which controls the energy balance in the interstellar space and the 
evolution of gas clouds. 
Also the surface of dust grains is an efficient site for formation of
H$_2$ molecules which act as an effective coolant at the time of
formation of stars from interstellar clouds, and enhance the star 
formation rate (SFR) and strongly affect the initial mass function (IMF)
in the metal--poor universe 
(Hirashita, Hunt, \& Ferrara 2002; Hirashita \& Ferrara 2002). 
In fact, 
the most iron--poor star HE0107--5240 so far discovered (Christlieb et
al. 2002) has raised important questions on the formation of such a
metal--poor ([Fe/H]=$-5.3 \pm 0.2$) but low--mass ($\sim$ 0.8
$M_{\odot}$) star.  Umeda \& Nomoto (2003a) showed that this star can be
the second generation star formed from a gas cloud enriched by a
population III supernova, which produced very little Fe but large enough
C and O for efficient gas cooling.  Schneider et al. (2003) also
considered this star as the second generation but argued the dust grains
produced by population III objects could play an important role in
forming low mass stars from such a metal--poor gas cloud as
[Fe/H]=$-5.1$.
Furthermore, dust grains residing in interstellar space in galaxies and
in intergalactic medium cause obscuration and reddening of starlight, 
and the thermal radiation from the dust grains distorts the 
cosmic background radiation (see Hauser \& Dwek 2001 for details). 
How much dust grains absorb stellar light and re--emit it by thermal
radiation heavily depends on their chemical composition, size and 
abundance. 
Therefore, the investigation of dust grains in the early universe is 
inevitable not only to reveal the structure and evolution of the early 
universe but also to deduce the SFR and the IMF during the evolution 
of the universe from the observations of the cosmic microwave background
(CMB) and the cosmic infrared background (CIB) which could be one of the
main subjects of the planned Atacama Large Millimeter Array (ALMA) and
the Next Generation Space Telescope (NGST) (Loeb \& Haiman 1997).

In cosmic environments, dust grains form in a cooling gas outflowing
from star to interstellar space such as in the stellar winds from AGB 
stars and in the ejecta of supernovae. 
The major source of dust grains in our Galaxy is considered to be AGB
stars evolving from stars with main--sequence mass $\le$ 8 $M_{\odot}$. 
However, the age of AGB stars is too old to contribute to dust grains in 
the early universe. 
Thus, supernovae evolving from  stars whose main--sequence mass is larger 
than 8 $M_{\odot}$ are considered to be the major source of dust grains 
in the early universe. 
Dust formation in the ejecta of supernovae has been suggested from the 
isotopic anomalies in meteorites, and the microscopic analysis of each 
dust grains extracted from meteorites has identified SiC, graphite and 
Si$_3$N$_4$ grains as the supernovae condensates (SUNOCONs) from their 
isotopic composition (see Zinner 1998 for details). 
SN 1987A is the first supernova in which the in--situ dust formation in 
the expanding ejecta was observed (Lucy et al. 1989; Whitelock et al.
1989; Meikle et al. 1993; Wooden et al. 1993; Colgan et al. 1994), and 
the model of formation of dust grains in the ejecta was investigated by 
Kozasa, Hasegawa, \& Nomoto (1989a,b, 1991). 
Also the dust formation was confirmed in the ejecta of SN 1999em from
the appearance of blue--shifted line emissions as observed in SN 1987A 
(Elmhamdi et al. 2003).

What kind of and how much dust grains condense in the ejecta have been 
still on debate. 
The observations of young supernova remnants (SNRs) with Infrared Space 
Observatory (ISO) revealed the thermal radiation from dust grains 
condensed in the ejecta of Cas A (Lagage et al. 1996; Arendt, Dwek,
 \& Moseley 1999), and the observed thermal radiation is well fitted 
by a mixture of Al$_2$O$_3$, MgSiO$_3$ and SiO$_2$ grains (Douvion,
Lagage, \& Pantin 2001), otherwise the spectroscopic observation at 
2.4--45 $\mu$m suggested that the pristine dust in Cas A is a peculiar 
class of silicate material (Arendt, Dwek, \& Moseley 1999). 
However, no thermal radiation originating from dust grains 
formed in the ejecta was observed towards the Tycho, Kepler and Crab SNRs 
by ISO (Douvion et al. 2001), despite that the optical observations have 
suggested the presence of dust grains in the Crab SNR (Fesen \& Blair 
1990; Hester et al. 1990). 
The estimated amount of hot dust observed in Cas A SNR by ISO (Arendt,
Dwek, \& Moseley 1999) is 10 to 100
times smaller than the previous value estimated from IRAS observations
(e.g., Mezger et al. 1986; Dwek et al. 1987). 
However the recent submillimeter observations by SCUBA have revealed the
existence of cold dust in the young SNR of Cas A and concluded 
that the amount of dust condensed in the ejecta is estimated to reach a few 
$M_{\odot}$ (Dunne et al. 2003).  
These observations strongly support that Type II supernovae could be the 
major source of dust grains in the early universe.

Of course, the chemical composition, size and amount of dust grains in 
the early universe are determined by the balance between production in 
the ejecta of supernovae and destruction by interstellar shock as well
as reverse shock penetrating into the ejecta (see, e.g., Dwek 1998
for details), and the investigation of dust formation in the ejecta 
of population III supernovae is the first step to reveal the nature of 
dust grains in the early universe. 
Todini \& Ferrara (2001) performed the dust formation calculations in 
the primordial core collapse supernovae with the progenitor mass $M_{\rm
pr}$ ranging from 12 to 35 $M_{\odot}$, adopting the models of supernovae 
by Woosley \& Weaver (1995) and applying the method of dust formation 
calculation for SN 1987A by Kozasa, Hasegawa, \& Nomoto (1989a, 1991) 
as a template. 
In the calculations, they considered one--zone model for elemental 
composition and density in the ejecta, assuming the uniform mixing of 
elements within the He--core. 
Taking account of formation and destruction of CO and SiO molecules,
they showed that carbon grain condenses first and then Al$_2$O$_3$, 
Mg--silicates and Fe$_3$O$_4$ grains condense in this order as the gas
cools down. 
Also, they showed that the size of newly formed grains is relatively 
small ($\le$ 300 $\rm{\AA}$ in radius) and is sensitive to the value of 
adiabatic index $\gamma$, and that the total mass of produced dust is 
about 0.08 to 0.3 $M_{\odot}$ per supernova.  

The recent theoretical investigations of fragmentation of gas clouds
in the metal--free early universe have claimed that the stars more
massive than some tens--100 $M_{\odot}$ populate the first generation 
stars (Nakamura \& Umemura 2001; Bromm, Coppi, \& Larson 2002). 
Such metal--free stars massive as 140 $M_{\odot} \le$
$M_{\rm pr} \le$ 260 $M_{\odot}$ may evolve stably to end up their lives
as pair--instability supernovae (e.g., Umeda \& Nomoto 2002; Heger \& Woosley 2002). 
Also Umeda \& Nomoto (2002) have suggested that hypernovae whose 
explosion energy is more than several to ten times that of the ordinary core 
collapse supernovae are necessary to reproduce the enhancement of Zn 
and iron--peak elements observed in metal--poor stars. 
Therefore, in this paper, we explore the dust formation in the ejecta 
of population III supernovae including hypernovae and pair--instability 
supernovae to investigate the dependence of the progenitor mass and 
the explosion energy on the formation of dust grains in the ejecta. 
Our main aim is to clarify the dependence of the yield of newly formed 
grains and their size on the progenitor mass, focusing on the ordinary 
core collapse supernovae and pair--instability supernovae. 
We also investigate the effects of the explosion energy and the mass of
$^{56}$Ni in the ejecta as well as the efficiency of formation of CO and 
SiO molecules on the formation of dust grains. 

As discussed by Kozasa, Hasegawa, \& Nomoto (1989a), the elemental
composition in the ejecta controls what kind of grain species condenses, 
and the temporal evolution of gas density and temperature in the ejecta 
strongly affects the size of newly formed grains in the ejecta.  
Thus, in the dust formation calculations we employ the hydrodynamical 
models and the elemental compositions of population III supernovae 
calculated by Umeda \& Nomoto (2002). 
The time evolution of gas temperature is calculated by the
multifrequency radiative transfer code taking account of the energy 
deposition from radio active elements (Iwamoto et al. 2000).  
The two extreme cases for the elemental composition in the ejecta 
are considered; 
the elemental composition with original onion--like structure 
(hereafter referred to as unmixed case) and the uniformly mixed case 
within the He--core. 
A theory of non--steady state nucleation and grain 
growth is applied for the calculation of dust grains under the
assumption that formation of CO and SiO molecules is complete. 
Formation of all possible condensates is taken into account 
simultaneously to clarify what kind of grain species really condenses. 

In section 2 we present the models of supernovae used in the calculations, 
and the method of dust formation calculation is described in section 3. 
The results of calculations for ordinary core collapse supernovae and 
pair--instability supernovae are presented and discussed in section 4, 
and summarized in section 5. 
The effects of the explosion energy and the amount of $^{56}$Ni in the 
ejecta as well as the efficiency of formation of CO and SiO molecules 
on the dust formation in the ejecta of supernovae are presented in 
Appendix A and B, respectively.

\section{Model of the population III supernovae}

The nature of population III supernovae has been extensively
investigated to decipher the chemical evolution of the early universe
compared with the observed elemental abundance of metal--poor stars
(Umeda \& Nomoto 2002, 2003b). 
Umeda \& Nomoto (2002) have suggested that, 
in addition to ordinary core 
collapse supernovae with the explosion energy $\sim$ $10^{51}$ erg, 
hypernovae whose explosion energy is larger than 5--50 $\times$
$10^{51}$ erg (e.g., Nomoto et al. 2001, 2003 for reviews) are necessary to
reproduce the enhancement of Zn and iron--peak elements observed in 
metal--poor stars. 
Also in the metal--free universe very massive stars with mass ranging 
from 140 to 260 $M_{\odot}$ evolve stably and explode as the 
pair--instability supernovae for which the star is completely disrupted 
by nuclear--powered explosion (Umeda \& Nomoto 2002; Heger \& Woosley 2002).

We apply the hydrodynamical models and the results of nucleosynthesis 
calculations of population III supernovae by Umeda \& Nomoto (2002) for 
the dust formation calculations.
The details of the model used in the present calculations are 
summarized in Table 1, where the labels C, P and H of the model 
represent ordinary core collapse supernovae (CCSNe), pair--instability 
supernovae (PISNe) and hypernovae (HNe), respectively, and the numerical 
value denotes the mass of progenitor in units of solar mass. 
The mass cut is the mass coordinate of the boundary between the ejecta 
and the remnant, and its value specifies the mass of $^{56}$Ni in the 
ejecta; $M(^{56}{\rm Ni})$ = 0.07 $M_{\odot}$ is taken as a typical
value for ordinary CCSNe to reproduce the observed behavior 
of early light curve for CCSNe. 
The models of HNe (H20A, B and H30A, B) are adopted to investigate the 
dependence of the explosion energy and the mass of $^{56}$Ni in the
ejecta on dust formation. 
The elemental composition and the temporal evolution of gas density 
and temperature in the ejecta, which are essential to investigate 
the dust formation, are summarized in the following subsections. 

\subsection{Elemental composition}

The elemental composition in the ejecta determines what kind of grain 
species condenses. 
No dust grain condenses in the hydrogen envelope of population III 
supernovae because the hydrogen envelope is metal--free. 
Even if heavy elements are intruded into the hydrogen envelope during
the evolution of progenitor, the low density and the high expansion 
velocity prevent dust grains from condensing in the hydrogen envelope as 
discussed by Kozasa, Hasegawa, \& Nomoto (1989a, 1991). 
Thus, the dust formation region in the ejecta of population III
supernovae is confined within the He--core where heavy elements exist. 

Figure 1 shows the elemental composition within the unmixed 
He--core at day 600 after the explosion, taking account of the decay 
of radio active elements; Fig. 1a for a model of CCSNe (C20) and 
Fig. 1b for PISNe (P170).
The ejecta is divided into five regimes according to the elemental 
composition of interest to dust formation; 
for example, in the ejecta of the model C20, 
Fe--Si--S layer (2.45 $M_{\odot} \le M_r \le$ 2.52 $M_{\odot}$), 
Si--S--Fe layer (2.52 $M_{\odot} \le M_r \le$ 2.95 $M_{\odot}$),
O--Si--Mg layer (2.95 $M_{\odot} \le M_r \le$ 3.14 $M_{\odot}$),
O--Mg--Si layer (3.14 $M_{\odot} \le M_r \le$ 4.92 $M_{\odot}$) and 
He--layer (4.92 $M_{\odot} \le M_r \le$ 5.79 $M_{\odot}$).

The He--layer is carbon--rich (C/O $>$ 1) for CCSNe and its
mass irregularly varies with the progenitor mass in the models
of CCSNe used in this paper. The mass of oxygen--rich layer (O--Si--Mg
layer and O--Mg--Si layer) increases with increasing the progenitor mass, 
and the mass of inner Si--S--Fe and Fe--Si--S layer depends on the value of 
the mass cut. On the other hand, in the ejecta of PISNe, 
the He--layer is very thin for P170 and the mass increases with
increasing the progenitor mass. 

The mixing of elements in the ejecta of SN 1987A was confirmed from the 
early emergence of X--rays and $\gamma$--rays (e.g., Dotani et
al. 1987; Itoh et al. 1987; Kumagai et al. 1988),
and also the hydrodynamical simulations (Hachisu et al. 1990; see 
Arnett et al. 1989 for a review and references) and the 
laboratory experiments (Drake et al. 2002) clearly demonstrate that the 
mixing of elements is caused by the Rayleigh--Taylor instability at the 
interface of each layer with the different elemental composition. 
The isotopic signatures in the presolar carbon and SiC grains identified 
as SUNOCONs have suggested the extensive and microscopic mixing of 
elements in the different layers in the explosion (Amari \& Zinner 
1997; Travaglio et al. 1999). Thus, as an ideal case,  the uniform
mixing of elements within the He--core was considered to investigate 
dust formation in the ejecta of supernovae (Kozasa, Hasegawa \& Nomoto 
1989a; Todini \& Ferrara 2001; Schneider, Ferrara, \& Salvaterra 2003).  
On the other hand, Douvion, Lagage, \& Pantin (2001) have suggested 
that the mixing in the ejecta of the Cas A SNR is knotty rather than 
microscopic. A simple calculation of molecular diffusion has shown that 
the mixing at atomic level is impossible before the condensation 
of dust grains in the ejecta (Deneault, Clayton, \& Heger 2003). 
It has been still on debate at this moment whether the mixing is 
macroscopic at knotty level or microscopic at atomic level and how large 
the mixing is extended in the ejecta. Therefore, in the following 
calculation of dust formation we consider the two extreme cases for the 
mixing of elements; one is the unmixed case with the original
onion--like structure,  and another is the mixed case for which the
elements are uniformly mixed within the He--core.  
 
\subsection{The time evolution of gas density and temperature in the
  ejecta}

The time evolution of gas density and temperature strongly affects  
the number density and size of newly formed grains in the ejecta. 
As the explosion shock propagates outwards, the reverse shock generated 
at the interface of layers with different chemical composition changes the
density structure. Although Deneault, Clayton, \& Heger (2003) show that 
the reverse shock generated at the He--H interface introduces a very
nonhomologous readjustment after $t=10^6$ seconds from the explosion,  
our hydrodynamical calculations including the interaction
with the reverse shock have shown that the interaction makes the 
expansion finally homologous in less than one day after the explosion 
(see Shigeyama \& Nomoto 1990).
Thus, the time evolution of gas density at a given mass coordinate
$M_r$ is calculated by 
\begin{eqnarray}
 \rho(M_r, t) = \rho(M_r, t_0)\left(\frac{t}{t_0}\right)^{-3} 
\end{eqnarray}
\noindent
where the reference time $t_0$ is one day after the
explosion.  
The distribution of gas density within
the He--core at day 600 after the explosion is shown in Figure 2 where
the mass coordinate is normalized to the He--core mass and the  
histogram presentation of the density structure reflects the size of mesh 
used in the calculations: Fig. 2a for
CCSNe (C20 and C25) and HNe (H25), and Fig. 2b for PISNe (P170 and P200). 
The gas density within the He--core of CCSNe with the same explosion 
energy is almost the same in the order of magnitude and does not depend 
so much on the progenitor mass, apart from the detailed structure caused 
by the propagation of reverse shock originating at the interface of each 
layer. 
The gas density in the ejecta of H25 is one order of magnitude lower
than that of CCSNe, reflecting the explosion energy of ten times of
CCSNe. 
Also the gas density in the ejecta of P200 is a little bit lower than
that in the ejecta of P170, but the gas density within the He--core is 
10$^{-14}$--10$^{-13}$ g cm$^{-3}$ at day 600 after the explosion and is 
almost the same order of the magnitude as the gas density in the ejecta 
of CCSNe, except for the region $M_r \le$ 40 $M_{\odot}$.

The gas temperature in the ejecta is determined by the detailed process
of the degradation of $\gamma$--rays and X--rays deposited from the
radio active elements. 
In the previous calculations (Kozasa, Hasegawa, \& Nomoto 1989a, 1991; 
Todini \& Ferrara 2001), the time evolution of gas temperature was 
approximated by 
\begin{eqnarray}
T(M_r, t) = T(M_r, t_0) \left(\frac{t}{t_0}\right)^{3(1-\gamma)}
\end{eqnarray}
\noindent
with a parameter $\gamma$, and the reference temperature $T(M_r, t_0)$ 
was determined by the observations of SN 1987A. However this method is 
not applicable for population III supernovae with no observational
data. 
Therefore, the time evolution of gas temperature is calculated by using 
the multifrequency radiative transfer code together with the energy 
equation, taking account of the deposition of energy from radio active 
elements (see Iwamoto et al. 2000 for details). 

As an example, Figs. 3a and 3b show the time evolution of gas
temperature at a location of oxygen--rich layer with almost the same 
elemental composition in the ejecta of CCSNe, PISNe and HNe; 
Fig. 3a for the unmixed ejecta of C20 (at $M_r=$ 3.7 $M_{\odot}$), 
C25 (at $M_r=$ 4.6 $M_{\odot}$), P170 (at $M_r=$ 56.5 $M_{\odot}$) and
P200 (at $M_r=$ 64.9 $M_{\odot}$), and Fig. 3b for the mixed ejecta of
C25 (at $M_r=$ 4.6 $M_{\odot}$) and P200 (at $M_r=$ 64.9 $M_{\odot}$) 
and for the unmixed ejecta of H30A (at $M_r=$ 6.8 $M_{\odot}$) and H30B 
(at $M_r=$ 6.8 $M_{\odot}$). 
As can be seen from Figure 3, it should be pointed out here that the 
temporal evolution of gas temperature in the ejecta is approximately 
determined only by the explosion energy and the amount of $^{56}$Ni in 
the ejecta and does not heavily depend on the progenitor mass in the 
mass range considered in this paper. 
Also the gas temperature in the mixed ejecta is almost the same as that
in the unmixed ejecta, except for the gas temperature around the day 200 
after the explosion. 
The gas temperature in the ejecta of H30 with the large explosion energy 
quickly decreases in comparison with CCSNe, because X--rays and 
$\gamma$--rays escape easily due to the low gas density in the ejecta. 
The effect of $M(^{56}{\rm Ni})$ in the ejecta on the gas temperature is 
clearly shown by the difference of the gas temperature in the unmixed 
ejecta between H30A and H30B. 
The gas temperature in the ejecta of PISNe with the large explosion 
energy is much higher than that in CCSNe because the progenitor is
massive and a large amount of $^{56}$Ni reside in the ejecta. 

From Figures 2 and 3, we can easily estimate the condensation time and 
the behavior of the average radius of a grain species formed in the
ejecta of population III supernovae; 
if the grains condense around gas temperature of 1500 K in the 
oxygen--rich layer, the condensation time $t_c$ after the explosion
would be around day 400 for CCSNe, day 600 for PISNe and day 200 for 
HNe with the same $M(^{56}{\rm Ni)}$ as CCSNe. 
The average radius of the condensate would be not so much different 
since the average size is roughly proportional to $\rho^{1/3} \propto 
(1/t_c)$, so far as the abundance of the elements available for dust 
formation is the same, which is confirmed by the results of dust
formation calculations given in section 4 and Appendix A. Note that 
the radiation transfer calculation does not include the radiative 
cooling of molecules such as CO and SiO which act as effective coolants 
of gas in the ejecta as demonstrated by Liu \& Dalgarno (1995) for SN
1987A. The effect of radiative cooling of molecules would make the 
condensation time of dust grains earlier and the average radius larger 
than those presented in section 4. However the average size and the 
number density of newly formed grains could not be so much different
so far as the condensation time is in the order of one hundred days. 

\section{Formulation and calculation of dust formation}

\subsection{Formulation of nucleation and grain growth}

In astrophysical environments, dust grains condense via formation of
condensation nuclei and their growth through the collisions of gaseous 
species in a cooling gas outflowing from star into interstellar space.

The nucleation rate is usually formulated under the condition of
presence of monomer molecule whose chemical composition is the same as 
the condensate. 
However, for formation of compound grains of astrophysical interest such 
as silicates, no gaseous monomer molecule exists and it is impossible to 
derive rigorously the nucleation rate without specifying the chemical 
passways and reaction constants.
Nevertheless, even if no information on chemical passways and reaction 
constants is available for formation of compound grains, as is discussed 
by Yamamoto et al. (2001), the nucleation rate can be evaluated by
applying the concept of the key species introduced by Kozasa \& Hasegawa 
(1987) when the net reaction rate is much larger than the decay rate. 
Under the assumption that the key species defined as the gaseous species 
of the least collision frequency among the reactants controls the
kinetics of nucleation and grain growth, the steady state nucleation
rate of $j$--th grain species $J^s_j(t)$ is given by 
\begin{eqnarray}
J_j^s(t) = \alpha_{sj} \Omega_j \left(\frac{2\sigma_j}{\pi
 m_{1j}}\right)^{1/2} \left(\frac{T}{T_d}\right)^{1/2} \Pi_j c_{1j} \exp 
\left[ -\frac{4}{27}\frac{\mu_j^3}{(\ln S_j)^2}\right] 
\end{eqnarray}
\noindent
where $\alpha_{sj}$ is the sticking probability of the key species for
$j$--th grain species, $\Omega_j$ is the volume of the condensate per
the key species, and $\sigma_j$ is the surface energy. 
We assume $\alpha_{sj}=1$ in the calculations (see Hasegawa \& Kozasa 
1988; Kozasa et al. 1996; Chigai, Yamamoto, \& Kozasa 1999 for 
the derivation of steady--state nucleation rate). 
The concentration and the mass of the key species for $j$--th grain
species are denoted by $c_{1j}$ and $m_{1j}$, respectively.  
$T_d$ is the temperature of the condensation nuclei and $T$ is the gas
temperature (see Kozasa et al. 1996). 
The factor $\Pi_j$ is a function of partial gas pressures of reactants 
and products except for the key species (Yamamoto et al. 2001) and we
put $\Pi_j=1$ in the calculations, partly because the appearance of the
factor is somehow related with the detailed chemical reaction mechanism 
at the condensation and partly because the factor can be negligible under
the condition that the number of the key species contained in the
condensation nuclei is much larger than unity. 
The quantity $\mu_j$ representing the energy barrier for nucleation is
defined by $\mu_j = 4 \pi a_{0j}^2 \sigma_j /kT_d$ with the hypothetical 
radius of condensate per the key species $a_{0j} = \left(3 
\Omega_j/4\pi\right)^{1/3}$, where $k$ is the Boltzmann constant. 
We assume the temperature of the condensation nuclei is the same as the 
gas temperature in what follows. 
The supersaturation ratio $S_j$ taking into account the chemical
reaction at the condensation of $j$--th grain species is calculated by 
\begin{eqnarray}
\ln S_j = -\frac{\Delta G_j^0}{kT} + \sum_i \nu_{ij} \ln P_{ij} 
\end{eqnarray}
\noindent
where the $\Delta G_j^0$ is the Gibbs free energy of formation of
$j$--th grain species from the reactants per the key species and 
$P_{ij}$ are the partial 
gas pressures of reactant and product gas species. 
The stoichiometric coefficients $\nu_{ij}$ of reactants and products are 
normalized to the key species; 
$\nu_{ij}$ is positive for a reactant and negative for a product
(see Kozasa \& Hasegawa 1987; Hasegawa \& Kozasa 1988). 

In the ejecta of supernovae where the gas cooling time and/or the 
collision time controlling the process of nucleation and grain growth
are comparable to the dynamical time scale, the application of the
steady state nucleation rate is questionable. 
In this paper, according to Gail, Keller, \& Sedlmayr (1984), we employ
the non--steady state nucleation rate $J_j(t)$ evaluated by the equation 
\begin{eqnarray}
\frac{\partial }{\partial t}\left(\frac{J_j(t)}{\eta_j}\right) = 
-\frac{1}{\tau_{*,j} \eta_j}\left[J_j(t)-J_j^s(t)\right]
\end{eqnarray}
\noindent 
with the relaxation time towards the steady state $\tau_{*,j}$ defined by 
\begin{eqnarray}
\tau_{*,j}^{-1} = 4 \pi a_{0j}^2 \alpha_{sj} \left(\frac{kT}{2 \pi m_{1j}}\right)^{1/2}
c_{1j}(t) \frac{\left(\ln S_j\right)^2}{\mu_j} = 
\tau_{{\rm coll},j}^{-1} \frac{\left(\ln S_j\right)^2}{\mu_j}
\end{eqnarray}
\noindent 
and $\eta_j = \tau_{{\rm coll},j}^{-1} (r_{c,j} / a_{0j})^2$, where 
$\tau_{{\rm coll},j}$ is the collision time of the key species and
$r_{c,j}=2 a_{0j} \mu_j/3 \ln S_j$ is the critical radius for $j$--th 
grain species. 

Given the nucleation rate $J_j(t)$ at a time $t$, the process of 
nucleation and grain growth is described by the two basic equations 
in a frame comoving with gas; 
one is the equation of continuity for the key species given by  
\begin{eqnarray}
1-\frac{c_{1j}(t)}{{\tilde c}_{1j}(t)}= 1- Y_{1j} = 
\int_{t_e}^{t}\frac{J_j(t')}{{\tilde c}_{1j}(t')}
   \frac{4\pi}{3\Omega_j}r_j^3(t,t')dt' \label{eqcons}, 
\end{eqnarray}
\noindent
where $t_e$ is the equilibrium time defined as a time at which 
the supersaturation ratio $S_j $ reaches to unity without depletion 
of the key species, and ${\tilde c}_{1j}$ is the nominal concentration 
of the key species expected without depletion of the key species due to 
nucleation and grain growth, and $Y_{1j} = c_{1j}/{\tilde
c}_{1j}$ represents the degree of depletion of the key species due to 
nucleation and grain growth, and $r_j(t,t')$ is the radius of grain
nucleated at $t'$ and measured at $t$. 
Another is the equation of grain growth given by 
\begin{eqnarray}
\frac{d r_j}{d t} = \alpha_{sj} \Omega_j
                  \left(\frac{kT}{2 \pi m_{1j}}\right)^{1/2}c_{1j}(t)
            = \frac{1}{3} a_{0j} \tau_{{\rm coll},j }^{-1}(t).
\end{eqnarray}
\noindent
By differentiating the equation (7) with $t$ subsequently, 
the integral equation is reduced to the simultaneous ordinary equations 
\begin{eqnarray}
\frac{dK^{(i)}_j}{dt} = \frac{J_j(t)}{{\tilde c}_{1j}(t)}\frac{4 \pi}{3 
\Omega_j} r_{c,j}^i + i K^{(i-1)}_j \frac{d r_j}{dt} ~ ~ 
({\rm for} ~i = 1-3)
\end{eqnarray}
\noindent
with
\begin{eqnarray}
\frac{dK^{(0)}_j}{dt} = \frac{J_j(t)}{{\tilde c}_{1j}(t)}\frac{4 \pi}{3 
\Omega_j}. 
\end{eqnarray}
\noindent
In principle, by solving the equations (5), (8), (9) and (10) in couple
with the temporal 
evolution of gas density $\rho(t)$ and temperature $T(t)$ with the
number abundances of reactants, the number density $n_{{\rm gr},j}(t)$ 
and volume equivalent average radius $ r_{{\rm gr},j}(t)$ of $j$--th 
grain species are calculated by 
\begin{eqnarray}
\frac{n_{{\rm gr},j}}{{\tilde c}_{1j}(t)} = \frac{K^{(0)}_j(t)}{a_{0j}^3} 
\end{eqnarray} 
\noindent
and 
\begin{eqnarray} 
r^3_{{\rm gr},j} = \frac{K^{(3)}_j(t)}{K^{(0)}_j(t)},
\end{eqnarray}
\noindent
respectively. Furthermore, the size distribution function $f_j(r)$ of 
newly formed $j$--th grain species at a time $t$ is calculated by 
\begin{eqnarray}
f_j(r)dr =\frac{{\tilde c}_{1j}(t)}{a_{0j}^3} \frac{d K^{(0)}_j(t')}{dt'}dt'
\end{eqnarray}
\noindent
since the grains with radii between $r$ and $r+dr$ are nucleated in 
the time interval of $t'$ and $t'+dt'$. 

\subsection{Calculation of dust formation in the ejecta of supernovae}

Formation of dust grains in the ejecta of supernovae is calculated by 
solving the equations (5), (8), (9) and (10) with the temporal evolution 
of gas density and temperature, given the grain species and the chemical 
reaction at the condensation with the abundance of the reactants.
The formation of CO and SiO molecules prior to the formation of dust 
grains was observed in the ejecta of SN 1987A (e.g., Bouchet \& Danziger
1993).  
Formation of CO and SiO molecules plays a crucial role on the formation 
of dust grains, because CO molecules lock the oxygen atoms available for 
formation of oxide grains and SiO molecules are considered to be the
starting molecules to form silicate grains. 

As discussed by Liu \& Dalgarno (1994, 1996), in the 
ejecta of supernovae, these molecules are destroyed by the impact with 
energetic electrons and charge transfer reactions with the ionized 
inert gaseous atoms which are created by the decay of radio active 
elements whose abundances depend on the elemental composition as well as 
the degree of the mixing in the ejecta. 
In this paper, to simplify the calculation of dust formation, we assume 
that formation of CO and SiO molecules is complete, that is, no 
carbon--bearing grain condenses in the region of C/O $<$ 1 and no 
Si--bearing grain except for the oxide grains condenses in the region of 
Si/O $<$ 1. The effect of incomplete formation of CO molecules has been
extensively investigated by Clayton, Liu, \& Dalgarno (1999) and
Clayton, Deneault, \& Meyer (2001) and they have shown that carbon
grain can condense even in the region of C/O $< 1$. Also the formation 
efficiency of SiO molecules is expected to affect the abundance and the
size of Si-- and/or Mg--bearing grains. The effect of the 
formation efficiency of CO and SiO molecules on the dust formation in 
the ejecta is discussed in Appendix B. 

The grain species, the chemical reactions at the condensations and the 
basic data necessary for dust formation calculations are tabulated in 
Table 2. 
The Gibbs free energy for formation of a condensate from the reactants 
is approximated by a formula $\Delta G^0_j/kT = -A/T +B $ and the 
numerical values of A and B are evaluated by the least squares' fitting 
of the thermodynamic data (Chase et al. 1985). 
It should be recognized here that what kind of grains species really 
condenses in the gas with a given elemental composition is determined by 
the competitive process of formation of each grain species, because 
formation of a grain species depletes the gaseous atoms and molecules 
available for the condensation of other grains. 
So in the calculations, formation of all possible condensates tabulated
in Table 2 is taken into account simultaneously.

\section{The results of calculations and discussions}

In this section, the results of calculations of dust formation in the 
ejecta of ordinary CCSNe and PISNe are presented and discussed. 
The behaviors of nucleation and growth of dust grains are depicted in 
section 4.1. 
The detailed results of the condensation times, the average radii, and 
the size distribution functions of each grain species formed in the
unmixed and mixed ejecta of CCSNe and PISNe are presented and discussed 
in sections 4.2 and 4.3, respectively. 
The dependence of the total mass of newly formed grains and that of the 
mass yield of each grain species on the progenitor mass are given in 
sections 4.4 and 4.5, respectively.  
The effects of the explosion energy as well as the amount of $^{56}$Ni
in the unmixed ejecta on dust formation are investigated in Appendix A, 
and the effect of the formation efficiency of CO and SiO molecules is 
discussed in Appendix B.

\subsection{The behavior of nucleation and grain growth} \label{bozomath0}

Figure 4 depicts the behaviors of nucleation and grain growth as a 
function of time after the explosion at a location of $M_r=$ 3.5 
$M_{\odot}$ in the O--Mg--Si layer with the elemental abundances 
relative to oxygen Si/O $ = 2.97 \times 10^{-2}$, Mg/O $ = 8.25 \times 
10^{-2}$, and Al/O $ = 9.38 \times 10^{-4}$;
Fig. 4a for the behaviors of nucleation rate $J_j(t)$ and the depletion 
of the key species $Y_{1j}$, and Fig. 4b for the number density 
$n_{{\rm gr},j}$ and the average radius $r_{{\rm gr},j}$. 
At this location, the expected condensates are Al$_2$O$_3$, Si-- and/or 
Mg--bearing grains. 
The key species is Al for Al$_2$O$_3$ and Mg or SiO for Mg-- and/or 
Si--bearing grains depending on what kind of grain species really 
condenses. 

As the gas cools down with time, it can be clearly shown from Figs. 4a 
and 4b that the nucleation rate rapidly increases with increasing the 
supersaturation ratio, and then decreases with the depletion of the key 
species due to grain growth, thus the nucleation rate has a maximum 
at a time. 
This is a typical behavior of nucleation and grain growth in a 
cooling gas. 
The time at which the nucleation rate reaches to the maximum is defined as 
the condensation time. 
At this location, Al$_2$O$_3$ grain condenses first at day 400, and then 
Mg$_2$SiO$_4$ condenses consuming the key species SiO at day 421, and 
finally the remaining Mg atoms are locked into MgO grains at day 439. 
Also it should be noted 
that each grain grows to the final radius in the very short time
interval; less than 20 days after the condensation time. 

Furthermore, we can see from Figs. 4a and 4b that the smaller abundance
of the key species leads to the larger number density of condensation
nuclei with broad peak of the nucleation rate and results in the small 
average radius. 
Thus, the abundance of the key species is very sensitive to the number 
density and average radius of newly formed grains, which is well 
reflected to the size distribution function of grains given in Figure 5;
Fig. 5a at $M_r=$ 3.5 $M_{\odot}$ and Fig. 5b at $M_r =$ 4.0 $M_{\odot}$ 
where the abundances of Si, Mg and Al are more than one order of 
magnitude smaller than at $M_r=$ 3.5 $M_{\odot}$; 
Si/O $ = 1.50 \times 10^{-4}$, Mg/O $ = 6.40 \times 
10^{-3}$, and Al/O $ = 3.95 \times 10^{-5}$.
At $M_r=$ 4.0 $M_{\odot}$, the smaller abundance of the key species SiO 
for formation of Mg$_2$SiO$_4$ grains results in the larger number of 
condensation nuclei with very broad peak of the nucleation rate
corresponding to the wide-spreaded size distribution, which is true for
 formation of Al$_2$O$_3$ grains. 
On the other hand, the size distribution function of MgO grains at 
$M_r =$ 4.0 $M_{\odot}$ is almost the same as that at $M_r=$ 3.5 
$M_{\odot}$, because the abundance of Mg being the key species for
formation of MgO and remaining after the formation of Mg$_2$SiO$_4$ is 
only factor 4 smaller at $M_r =$ 4 $M_{\odot}$ than at $M_r =$ 3.5 
$M_{\odot}$.  
Note that the size distribution function of a grain species formed at a 
location is approximately log--normal so far as the average size is 
larger than $\sim$ 0.01 $\mu$m.

\subsection{The dust formation in unmixed ejecta of CCSNe and PISNe} \label{bozomath1}

In the unmixed ejecta of supernovae, a variety of grain species condense 
in each layer corresponding to the difference in the elemental composition. 
Fig. 6a and Fig. 6b show the condensation times of dust grains formed in 
the unmixed ejecta of C20 and P170, respectively.
At first carbon grains condense in the He--layer, which is followed by 
the condensation of Al$_2$O$_3$ and Mg--silicates (Mg$_2$SiO$_4$ and
MgSiO$_3$) in the oxygen--rich layer, MgO in the O--Mg--Si layer,
SiO$_2$ in O--Si--Mg layer, Si and FeS inside Si--S--Fe layer, and Fe in 
the innermost Fe--Si--S layer in this sequence. 
Note that in the O--Si--Mg layer SiO$_2$ grains condense from SiO
molecules left over after the formation of Mg--silicate grains,
otherwise MgO grains condense from Mg atoms remaining in the O--Mg--Si
layer.
The major grain species are Mg$_2$SiO$_4$ and MgO in the O--Mg--Si
layer, otherwise MgSiO$_3$ and SiO$_2$ in the O--Si--Mg layer. 
In the ejecta of C20, Si grains condense in the innermost region of the 
He--layer because a significant amount of Si atoms are present around
the interface between He--layer and oxygen--rich layer (see Fig. 1a),
but no SiC grain condenses in this region. 
Anyway, in the unmixed ejecta of C20 and P170, the grain species
condensed in the ejecta are the same. 
Also metallic Cr and Ni grains do not condense significantly in the 
Fe--Si--S layer of the both ejecta because of the low number density 
as well as the high energy barrier for nucleation.   

The condensation time generally depends not only on the temporal
evolution of gas temperature but also on the concentration of the key 
species at the formation site. 
On the other hand, given the time evolution of gas temperature, the 
condensation temperature defined as the gas temperature at the
condensation time depends only on the concentration of the key species; 
the condensation temperature decreases with decreasing the
concentrations. 
In the unmixed ejecta of CCSNe and PISNe, the condensation temperature
does not heavily depend on the progenitor mass; 
$\sim$ 1900 K for carbon, 1600--1700 K for Al$_2$O$_3$, 1400--1500 K for 
Mg--silicates, 1350--1450 K for MgO, 1300--1400 K for SiO$_2$,
1100--1200 K for Si, 1000--1100 K for FeS and 800--850 K for Fe. 
The reason is as follows; as described in section 2, the concentration of
the key species is a few times large in the ejecta of PISNe compared with
CCSNe, and the elemental composition in each layer and the gas density
at a time are not so significantly different between CCSNe and PISNe. 
On the other hand, the gas temperature in the ejecta of PISNe is higher 
than that of CCSNe. 
Thus, as can be seen from Figs. 6a and 6b, the condensation time of each 
grain species in the ejecta of PISNe is about 150 days delayed in 
comparison with that of CCSNe. 
Anyway, dust grains condense in the ejecta of ordinary CCSNe around day 
300 to 600 after the explosion, and around day 500 to 800 in the ejecta
of PISNe, which is almost independent of the progenitor mass. 

Figures 7a and 7b show the average radius of each grain species in the 
ejecta of C20 and P170, respectively. 
The average radius of each grain species well reflects the concentration
of the key species at the condensation time, and heavily depends on the
elemental composition and the gas density at the formation
site in the ejecta. 
In the ejecta of C20, the average radii of Fe and Si condensed in the 
innermost region are relatively large; 
about 0.2 $\mu$m for Fe and 0.4 $\mu$m for Si. 
The range of the average radius spans about one order of magnitude for 
carbon, Al$_2$O$_3$, SiO$_2$ and FeS whose maximum radii are 0.4 $\mu$m, 
0.004 $\mu$m, 0.1 $\mu$m and 0.06 $\mu$m, respectively. 
The range spans more than two orders of magnitude for MgO, 
Mg$_2$SiO$_4$, and MgSiO$_3$ with the maximum radii 0.2 $\mu$m, 0.1 $\mu$m 
and 0.5 $\mu$m, respectively. 
In the ejecta of P170, the average radius of each grain species is a
little smaller than that in the ejecta of C20, but the difference is not 
significant except for Fe, FeS and Si condensed in the Fe--Si--S layer 
where the gas density is substantially lower than that
in the ejecta of C20 (see Fig. 2). 
The average radii of dust grains formed in the unmixed ejecta are
limited to less than 1 $\mu$m. 

The size distrubution functions of newly formed grains in the ejecta of 
C20 and P170 are given in Figs. 8a and 8b, respectively. 
As is shown in $\S$ 4.1, the size distribution function at a location
is approximately log--nomal for a grain species with the average radius 
larger than $\sim$ 0.01 $\mu$m. 
However, the size distribution function of a grain species summed up
over the formation region does not keep the original shape at the
formation site, reflecting the difference of the elemental composition
and the gas density in the formation region. 
Generally, for a grain species whose formation region is wide--spreaded 
in the ejecta and whose average radius spans more than one order of 
magnitude, the size distribution function completely deviates from 
the original log--normal shape, which is true for MgO, 
Mg$_2$SiO$_4$, MgSiO$_3$ and FeS grains. 
Although the average size of carbon grains ranges from 0.01 to 0.4
$\mu$m, the size distribution function looks like log--normal because
the large grains formed in the innermost region of He--layer do not 
contribute to the size distribution function due to the small number 
compared with the small grains condensed in the outer He--layer. 
The size distribution functions of SiO$_2$ and Fe in the ejecta are 
approximately log--normal. 
Si grain shows a bimordial distribution function with large grains 
condensed inside the Si--S--Fe layer and smaller ones condensed around
the interface between the He--layer and the O--Mg--Si layer. 
In the ejecta of P170, the behavior of the summed up size distribution 
function of each grain species is the same as that in the ejecta of
C20. 
However, in the ejecta of P170, the bimordiality of Si stems from the 
wide span of the average radius arising from the density variation 
within the Fe--Si--S layer. 

The size distribtion function summed up over all grain species formed 
within the He--core is drawn by the thick curve in Figure 8. 
The thick straight lines represent power--law formulae with the index
of $\alpha = -3.5$ and $\alpha = -2.5$. 
As can be seen from Fig. 8, in the both cases, the size distribution 
summed up over all grain species is well fitted with a power--law formula
whose index is $-3.5$ for the larger size and $-2.5$ for smaller one,
which is different from the conventional MRN size distribution function 
(Mathis, Rumpl, \& Nordsieck 1977) used in the astrophysical literatures. 
The radius at the crossover point is 0.06 $\mu$m for C20 and 0.02 
$\mu$m for P170. 
Although in this subsection only one model for CCSNe and PISNe is 
presented, respectively, the behaviors of dust formation and the average 
size as well as the size distribution function described above are
almost the same for the other models, being not dependent on the 
progenitor mass. 

\subsection{The dust formation in uniformly mixed ejecta of CCSNe and
  PISNe} \label{bozomath2}

This subsection presents the results of the dust formation calculations 
in the ejecta with the elemental composition uniformly mixed within the 
He--core as a extreme case. 
In the calculations we assume the density structure in the ejecta is not
affected by the mixing, since the numerical simulation suggested that 
the Rayleigh--Taylor instability produces the clumpy structure but the 
gas density does not so much change on average (Hachisu et al. 1990).
 
Under the assumption that formation of CO molecules is complete, only 
oxide grains condense in the uniformly mixed ejecta because the ejecta
is oxygen--rich. 
Figure 9 shows the condensation time of each grain species formed in the 
ejecta; Fig. 9a for C25 and Fig. 9b for P200. 
In the both models, Al$_2$O$_3$, Mg--silicates (Mg$_2$SiO$_4$ and
MgSiO$_3$), SiO$_2$ and Fe$_3$O$_4$ grains condense in this order. 
Being different from the unmixed ejecta, no MgO grain condenses
because of Si $>$ Mg in the uniformly mixed ejecta, and all iron atoms 
are locked into Fe$_3$O$_4$ grains. 
Except for the inner-- and outermost region of P200, the condensation 
temperature of each dust grains is almost the same within the He--core; 
1550--1600 K for Al$_2$O$_3$, 1450 K for Mg--silicates, 1350 K for
SiO$_2$ and 1300--1350 K for Fe$_3$O$_4$. 

In the ejecta of C25, the gas temperature decreases faster in the outer 
region than in the inner region of $M_r \le$ 6.8 $M_{\odot}$ where the 
gas temperature at a time $t$ is almost the same.  
Thus, well reflecting the temperature structure in the ejecta, in the 
ejecta of C25, Al$_2$O$_3$ grains start to condense at day 330 in the 
outer edge of He--core after the explosion, Mg--silicate grains at day
350, SiO$_2$ grains at day 364, and Fe$_3$O$_4$ at day 370. 
In the region of $M_r \le$ 6.8 $M_{\odot}$, each grain species condenses 
at almost the same time; 
Al$_2$O$_3$ around day 385, Mg--silicates around day 400, 
SiO$_2$ around day 415 and Fe$_3$O$_4$ around day 420. 
Being different from the formation of dust grains in the unmixed ejecta, 
all dust grains condense within 100 days after the onset of dust formation.
On the other hand, in the ejecta of P200, the condensation time of each 
grain species increases with decreasing the mass coordinate because the 
gas density gradually decreases with decreasing $M_r$;
Al$_2$O$_3$ grains from day 450 up to day 560, Mg--silicate grains from 
day 450 up to 570, SiO$_2$ grains from day 465 up to day 583, and
Fe$_3$O$_4$ grains from day 470 to day 590. 
In the ejecta of P200, the condensation times of dust grains are about
150 days later than those in the ejecta of C25 since the gas temperature 
at a given time is higher than that of C25. 

As are shown in Fig. 10a for C25 and Fig. 10b for P200, the range of
average radius of each grain species spans less than one order of
magnitude throughout the He--core. 
Given the same elemental composition within the He--core, the average 
radius depends on not only the density structure of gas but also the 
time evolution of gas temperature; in the outer region of the He--core, 
the gas density is lower than that in the inner region, but the gas 
temperature decreases faster in the outer region, which compensates the 
decrease of gas density at the condensation time. 
In the inner region of P200, the gas temperature is almost the same
being independent of $M_r$ and the variation of the gas density is less 
than one order of magnitude.
In the mixed ejecta of C25, the average radii of SiO$_2$ and 
Mg--silicates are a few hundredth $\mu$m with the maximum 0.1 $\mu$m 
for SiO$_2$, 0.05 $\mu$m for Mg$_2$SiO$_4$ and 0.07 $\mu$m for
MgSiO$_3$. 
The average radii of Fe$_3$O$_4$ and Al$_2$O$_3$ are small; several tens 
${\rm \AA}$ for  Fe$_3$O$_4$ and a several ${\rm AA}$ for Al$_2$O$_3$
which are consistent with the results by Todini and Ferrara (2001). 
Despite that the gas temperature in the mixed ejecta of P200 is higher 
than that of C25, the average radius of each grain species formed in the 
region of $M_r >$ 40 $M_{\odot}$ is almost the same as that in the
ejecta of C25, because abundance of the key species of dust grains is a 
few times larger in P200 than in C25. The low density in the region of 
$M_r <$ 40 $M_{\odot}$ of P200 (see Fig. 2b) results in the smaller 
average radius for each grain species.

Figure 11 shows the size distribution function of each dust grains formed
in the mixed ejecta; Fig. 11a for C25 and Fig. 11b for P200. 
Except for Al$_2$O$_3$ grains with very small average radius, in the
both models, the size distribution functions of Mg$_2$SiO$_4$,
MgSiO$_3$, SiO$_2$ and Fe$_3$O$_4$ grains approximately tend to be 
log--normal because of the narrow size range of each grain species. 
The size distribution function summed up over all grain species is 
also fitted with a power--law formula, whose index is $-3.5$ for
radius larger than 0.004 $\mu$m in P200 and $-2.5$ for radius smaller 
than 0.02 $\mu$m in C25. 
The deviation from the power--law formula is distinctive for the larger 
size in the model of C25, because only SiO$_2$ grain with the
log--normal distribution function contributes to the larger size. 
However, this is an untypical case for the behavior of the summed up
size distribution function. 
Here it should be addressed that, irrespective of the progenitor mass, 
the size distribution function summed up over all grain species newly 
formed in the ejecta with and without mixing is approximated by
power--law distribution whose index is $-3.5$ for larger radius and 
$-2.5$ for smaller one, although the radius at the crossover point
depends on the model of supernovae with the range from 0.004 to 0.1 $\mu$m. 

\subsection{The amount of freshly formed dust grains} \label{bozomath3}

Figure 12 shows the total mass of dust grains produced in the ejecta 
versus the progenitor mass; Fig. 12a for CCSNe and Fig. 12b for PISNe, 
where open triangle denotes the results of calculations for the unmixed 
ejecta and open square for the mixed ejecta. 
In Fig. 12a, for a reference, the results of calculations in the unmixed 
ejecta of HNe are also plotted by cross for H25A and H30A, and open
circle for H25B and H30B. 
The straight lines in Fig. 12a and in Fig. 12b are the linear least 
squares' fits to the calculated mass for ordinary CCSNe and linearly
connect the two data points for PISNe, respectively; solid line for the 
unmixed ejecta and dashed line for the mixed ejecta. 

The total mass of newly formed grains increases with increasing the
progenitor mass in the both unmixed and mixed ejecta of CCSNe and PISNe. 
In the unmixed ejecta of ordinary CCSNe, the total dust mass is 0.57 
$M_{\odot}$ for C20 and 1.32 $M_{\odot}$ for C30. 
It is generally expected that the amount of dust grains is more in the 
mixed ejecta than in the unmixed ejecta because much more oxygen atoms 
are locked into dust grains in the mixed ejecta. 
However, in the mixed ejecta of C13, the total mass of dust grains is 
0.22 $M_{\odot}$ and is a little smaller than the dust mass 0.23 
$M_{\odot}$ in the unmixed ejecta. 
The reason is that in the mixed ejecta of C13, formation of CO molecules 
consuming the oxygen atoms available for oxide grains limits the amount
of oxide grains; in this case, a small amount of Si and FeS grains
condense instead of SiO$_2$ and Fe$_3$O$_4$ grains. 
Also in the model of C25, the dust mass in the mixed ejecta is not so 
much different compared with that in the unmixed ejecta, because in the 
model of C25 the He--layer massive compared with other CCSNe somehow 
depresses the increase in the amount of mass available for formation of 
oxide grains within the oxygen--rich layer. 

In the unmixed ejecta of hypernovae with the same $M(^{56}{\rm Ni})=$ 
0.07 $M_{\odot}$ as CCSNe, H25A and H30A produce almost the same dust 
mass as C25 and C30, respectively. 
On the other hand, with a deep mass cut corresponding to 
$M(^{56}{\rm Ni}) =$ 0.7 $M_{\odot}$, H25B and H30B produce the dust 
grains about 0.7 $M_{\odot}$ larger than C25 and C30, respectively.
In the ejecta of the CCSNe except for the model of H25 and H30, the
ratio of dust mass to the progenitor mass is 0.02--0.05, which increases 
with increasing the progenitor mass. 
Although the total mass of newly formed grains depends on the detailed 
elemental composition in the ejecta as well as the position of mass cut, 
in general, the total mass produced in the mixed ejecta becomes larger 
than that in the unmixed ejecta with increasing $M_{\rm pr}$.  

PISNe, for which the progenitor is very massive and is completely 
disrupted, produce much more dust grains than CCSNe. 
For the model P200, the total mass of newly formed grains is $\sim$ 40 
$M_{\odot}$ in the unmixed ejecta and $\sim$ 60 $M_{\odot}$ in the mixed 
ejecta. 
The mass ratio of newly formed grains to the progenitor mass for PISNe
reaches to 0.15--0.3 at least, and the ratio increases with increasing
the progenitor mass.  
Therefore, in the early universe, much more dust grains could be 
produced and be injected into interstellar space if the very massive 
stars populate the first generation stars. 

\subsection{The mass yield of each grain species}

Figures 13 and 14 present the mass yield of each grain species versus 
the progenitor mass $M_{\rm pr}$ in the unmixed ejecta and the mixed 
ejecta, respectively; Figs. 13a and 14a for CCSNe, and Figs. 13b and 14b 
for PISNe. 
In the figures, the smooth curves are the least squares' spline fits to 
the calculated values for CCSNe and are the straight lines connecting the two 
calculated values in linear scale for PISNe.
Thus for PISNe, the mass yield for a given progenitor mass should be 
taken as a nominal value.

In the unmixed ejecta of CCSNe (Fig. 13a), with increasing $M_{\rm pr}$, 
the mass yield of carbon grains, apart from the small fluctuation,
somehow increases up to around $M_{\rm pr} =$ 25 $M_{\odot}$, and then 
decreases, while the yields of other grain species except for Fe grain
increase. 
The grain species contributing to the total dust mass heavily depend on 
the progenitor mass for $M_{\rm pr} \le$ 28 $M_{\odot}$;
in the ejecta with $M_{\rm pr} <$ 15 $M_{\odot}$, the major grain species 
in mass are carbon, Fe and Si grains with the average radius $\ge$ 0.1 
$\mu$m. 
With increasing $M_{\rm pr}$, the masses of the oxygen--rich layer as 
well as of the Si--S--Fe layer increase. 
Therefore, the yields of Si and Mg$_2$SiO$_4$ increase, and also the
mass of FeS increases together with decreasing the mass of Fe. 
Around $M_{\rm pr} =$ 20 $M_{\odot}$, Mg$_2$SiO$_4$, Si and FeS grains 
tend to dominate the mass abundance. 
Around $M_{\rm pr} =$ 25 $M_{\odot}$, the major grain species are 
Mg$_2$SiO$_4$, Si and C grains. 
The mass abundances of FeS, SiO$_2$ and MgO grains are larger than that 
of carbon grain for $M_{\rm pr} \ge$ 28 $M_{\odot}$, and Mg$_2$SiO$_4$ 
and Si are the major grain species followed by MgO, SiO$_2$ and FeS grains. 
Although the masses of Al$_2$O$_3$ and MgSiO$_3$ grains increase with $M_{\rm{
pr}}$, their mass fractions are limited to less than 10$^{-2}$. 

The masses of Mg$_2$SiO$_4$ and MgO grains per supernova are almost
constant in the unmixed ejecta of PISNe, being almost independent of the 
progenitor mass (see Fig. 13b); 
$M$(Mg$_2$SiO$_4$) $\simeq$ 6 $M_{\odot}$ and slightly decreases with 
increasing $M_{\rm pr}$, and $M$(MgO) $=$ 1.4 $M_{\odot}$. 
When the progenitor mass is smaller than 160 $M_{\odot}$, Mg$_2$SiO$_4$, 
Si, and SiO$_2$ grains are the major grain species contributing to the 
total mass. 
In contrast to the unmixed ejecta of CCSNe, the mass fraction of carbon 
grain is to be less than 5 $\times$ 10$^{-2}$ although the mass of
carbon grains increases with increasing $M_{\rm pr}$. 
It should be noted here that for 
$M_{\rm pr} >$ 160 $M_{\odot}$, the most abundant grain
species in the unmixed ejecta of PISNe is Si grain followed by SiO$_2$
grains, because the masses of the Si--S--Fe layer and the O--Si--Mg layer 
increase with increasing $M_{\rm pr}$. 
Also the mass fractions of iron--bearing grains (Fe and FeS) produced
within the Si--S--Fe layer are larger than that of 
Mg$_2$SiO$_4$ grains for $M_{\rm pr} >$ 220 $M_{\odot}$. 

In the mixed ejecta of CCSNe (see Fig. 14a), SiO$_2$ and Mg$_2$SiO$_4$ 
grains mainly contribute to the total dust mass for $M_{\rm pr} \le$ 15
$M_{\odot}$. 
The mass of Fe$_3$O$_4$ for $M_{\rm pr} \ge$ 20 $M_{\odot}$ is $\sim$ 
0.09 $M_{\odot}$, being almost independent of the progenitor mass. 
With increasing $M_{\rm pr}$, the mass of Mg$_2$SiO$_4$ increases and
reaches to the same as that of SiO$_2$ at $M_{\rm pr} =$ 30
$M_{\odot}$. 
The dust mass of massive CCSNe with $M_{\rm pr} \ge$ 25 $M_{\odot}$ is 
dominated by SiO$_2$ and Mg$_2$SiO$_4$ grains. 
Although the yield of Al$_2$O$_3$ increases with increasing $M_{\rm pr}$ 
and tends to be the same mass as Fe$_3$O$_4$, the mass fraction of 
Fe$_3$O$_4$ is limited to less than 0.1 for $M_{\rm pr} \ge$ 20
$M_{\odot}$ and that of Al$_2$O$_3$ is limited to less than 0.02 at
most. 
In the mixed ejecta of PISNe (see Fig. 14b), SiO$_2$ is the most
abundant grain, regardless of the progenitor mass. 
As is the same as the mixed ejecta of CCSNe, around the low--mass end of 
the progenitor mass of PISNe, SiO$_2$ and Mg$_2$SiO$_4$ contribute to the 
total mass. 
As the progenitor mass increases, the masses of MgSiO$_3$ and Fe$_3$O$_4$
increase and approach to the almost the same mass, while the amount of 
Mg$_2$SiO$_4$ decreases. 
The total mass of dust grains is dominated by SiO$_2$, MgSiO$_3$ and
Fe$_3$O$_4$ whose mass fractions are $\sim$ 0.7, 0.15 and 0.15, 
respectively.

\section{Summary}

We investigated the dust formation in the ejecta of population III
core collapse and pair--instability supernovae, employing the
hydrodynamical models and the results of nucleosynthesis calculations
by Umeda and Nomoto (2002). 
The time evolution of gas temperature in the ejecta is calculated by 
solving the radiative transfer equation together with the energy
equation taking into account of the energy deposition from radio active 
elements. 
Two extreme cases are considered for the elemental composition within 
the He--core; the unmixed case and the uniformly mixed case. 
In order to clarify what kind of grain species condenses in the ejecta, 
formation of all possible condensates is calculated simultaneously by 
applying a theory of non--steady state nucleation and grain growth under 
the assumption of complete formation of CO and SiO molecules. 
The results of the calculations are summarized as follows; 

 (1) The species of newly formed dust grains are affected by the
 elemental composition in the ejecta.  
 In the unmixed ejecta, a variety of grain species condense
 corresponding to the elemental composition at the formation site;  C 
 and Si grains in the He--layer, Al$_2$O$_3$, Mg$_2$SiO$_4$ and MgSiO$_3$
 grains in the oxygen--rich layer (O--Mg--Si and O--Si--Mg layers), MgO
 grain in the O--Mg--Si layer, SiO$_2$ grain in the O--Si--Mg layer, Si
 and FeS grains in the Si--S--Fe layer and Fe grain in the Fe--Si--S
 layer. 
 On the other hand, only oxide grains such as Al$_2$O$_3$,
 Mg$_2$SiO$_4$, MgSiO$_3$, SiO$_2$ and Fe$_3$O$_4$ condense in the mixed 
 ejecta. 
 The main species of newly formed grains do not depend on the progenitor 
 mass.

 (2) The average size of newly formed dust grains strongly depends on 
 the concentration of the key species at the condensation time as well
 as at the formation site. 
 In the unmixed ejecta, the range of average radius of each grain
 species spans a few orders of magnitude, depending on the grain species 
 and the formation region. 
 On the other hand, in the mixed ejecta, the average radius of each
 grain species is almost the same order of magnitude in the entire
 region within the He--core, reflecting the uniform  elemental
 composition. 
 The average radius of each grain species in both the unmixed
 and the mixed ejecta does not strongly depend on the progenitor mass
 and the explosion energy, but is affected by the amount of $^{56}$Ni in
 the ejecta. 
 No dust grain condenses with the average radius larger than 1 $\mu$m in
 the ejecta of population III supernovae considered in this paper.
  
 (3) The size distribution function of a grain species condensed at a 
 location in the ejecta is almost log--normal so far as the average
 radius is larger than 0.01 $\mu$m. 
 The size distribution function summed up within the He--core is log--normal 
 for carbon, Fe and SiO$_2$ grains condensed in the narrow confined 
 region of the unmixed ejecta, otherwise in the mixed ejecta, except for 
 Al$_2$O$_3$ grain, the summed up size distribution function is
 log--normal. 
 The size distribution function summed up over all grain species is 
 approximated by a power--law formula with the index of $-3.5$ for the
 larger radius and $-2.5$ for the smaller one, which is true for the
 unmixed and the mixed cases, being almost independent of the progenitor 
 mass; the radius at the crossover point ranges from 0.004 to 0.1 $\mu$m, 
 depending on the model of supernovae.

 (4) The total mass of newly formed dust grains increases with
 increasing the progenitor mass and is generally much more in the mixed
 case than in the unmixed case. 
 The total mass of dust grains is 2--5 \% of the progenitor mass for 
 CCSNe with 13 $M_{\odot} \le$ $M_{\rm{pr}} \le$ 40 $M_{\odot}$, and 
 15--30 \% of the progenitor mass for PISNe with 140 $M_{\odot} \le$ 
 $M_{\rm{pr}} \le$ 260 $M_{\odot}$.\footnote{After submitting this 
paper to ApJ on 28 June 2003, we have 
received a preprint from Dr. A. Ferrara, who obtained basically
similar results for the mass of dust produced per PISN as ours.}
Thus, if PISNe populate the early universe, a large amount of dust
grains would be injected into the primordial interstellar medium. 

 (5) The mass yield of each grain species depends on the progenitor mass. 
 C, Fe and Si grains are the major grains species in the unmixed ejecta
 of CCSNe with $M_{\rm pr} <$ 15 $M_{\odot}$. 
 With increasing $M_{\rm pr}$, Si and Mg$_2$SiO$_4$ grains are getting 
 dominant, and Mg$_2$SiO$_4$ grains is the most abundant for $M_{\rm pr}
 \ge$ 24 $M_{\odot}$. 
 In the unmixed ejecta of PISNe, the most abundant grain is
 Mg$_2$SiO$_4$ for $M_{\rm pr} <$ 160 $M_{\odot}$, otherwise for $M_{\rm
 pr} \ge$ 160 $M_{\odot}$ the most abundant grain is Si followed by
 Mg$_2$SiO$_4$, SiO$_2$, FeS, and Fe grains. 
 In the mixed ejecta of CCSNe, SiO$_2$ and/or Mg$_2$SiO$_4$ grains are 
 the major grain species contributing to the total mass. 
 Being independent of the progenitor mass, in the mixed ejecta of
 PISNe, SiO$_2$ grain is the most abundant species, which is followed by 
 Mg$_2$SiO$_4$ grains for $M_{\rm pr} <$ 160 $M_{\odot}$ and by
 MgSiO$_3$ and Fe$_3$O$_4$ grains for $M_{\rm pr} \ge$ 170 $M_{\odot}$. 

In the calculations, we treat only the formation of homogeneous grains
taking into account of the chemical reaction at the condensation but do 
not consider the possibility for the heterogeneous nucleation on the 
surface of pre--condensed grains. 
In fact, the heterogeneous grains consisting of metal--carbide cores and 
a graphite mantle have been observed in the Murchison meteorite 
(Bernatowicz et al. 1991, 1996). 
Also we must consider the possibility of the formation of composite 
grains such as Fe--Ni--Cr alloy to reveal whether the metal elements of 
Ni, Co and Cr with relatively small abundance are locked into dust
grains in the ejecta of supernovae or not. 
Also as is discussed in Appendix B, it is very important to investigate 
the effect of the formation efficiency of CO and SiO molecules on the 
formation of dust grains in order to clarify what kind of dust grains 
really condenses in the ejecta of supernovae. 
These subjects are left for the future work. 

Anyway, dust grains in the early universe play a critical role on formation
and evolution history of stars and galaxies. 
The investigation of the nature of dust grains in the early universe is 
essential not only to investigate the evolution and the structure of the 
early universe but also to deduce the SFR and the IMF during the
evolution of the universe from the relevant observations, because how
much dust grains absorb stellar light and re--emit it by thermal
radiation depends on the chemical composition, the size distribution and 
the amount of dust grains residing in interstellar space in galaxies and 
in intergalactic medium. 
The evolution of dust grains is determined by the balance between the
production in the ejecta of supernovae and the destruction by the 
interstellar shock as well as the reverse shock penetrating into the
ejecta. 
The efficiency of destruction by shock depends on the chemical
composition and the size of dust grains. 
Therefore, the results of the calculations presented in this paper can
be used as the basis to investigate the evolution of dust grains in the 
early universe. 

\acknowledgments
The authors thank the anonymous referee for 
the critical comments which improved the manuscript. 
This work has been supported in part by the Grant-in-Aid for Scientific
Research (13640229) from the Japan Society for the Promotion of Science.

\appendix

\section{The dust formation in the ejecta of hypernovae}

The hypernovae with the explosion energy larger than 10$^{52}$ erg are
proposed to reproduce the behavior of observed early light curve of SN
1998bw (Iwamoto et al. 1998; Woosley, Eastman, \& Schmidt 1999). 
Also the analysis of the light curve of SN 1998bw (Nakamura et al. 2001) 
concluded that the mass of $^{56}$Ni in the ejecta is 0.4 $M_{\odot}$, 
which is larger than a typical value of 0.07 $M_{\odot}$ in the ejecta
of ordinary CCSNe. 
Furthermore, Umeda and Nomoto (2002) have claimed that the primordial 
hypernovae well reproduce the observed elemental composition in the 
metal--poor stars; [Zn/Fe] increasing with decreasing [Fe/H] for [Fe/H] 
$\le -2.5$ (Primas et al. 2000).  
In this Appendix, we shall perform the calculations of dust formation in 
the ejecta of population III hypernovae to investigate the effect of the 
explosion energy as well as the amount of $^{56}$Ni in the ejecta on the
 dust formation. 
We consider two model; one is H30A with $M_{\rm pr} =$ 30 $M_{\odot}$, 
$E_{\rm exp} = 3 \times 10^{52}$ erg and $M(^{56}{\rm Ni}) =$ 0.07 
$M_{\odot}$, and another is H30B for which $M_{\rm pr}$ and $E_{\rm
exp}$ are the same as H30A but  $M(^{56}{\rm Ni}) =$ 0.7 $M_{\odot}$.

Fig. 15a shows the condensation time of each dust grains in the unmixed 
ejecta of H30A with the explosion energy 30 times but $M(^{56}{\rm
Ni})$ as same as ordinary CCSNe. 
The grain species formed in the ejecta and their condensation sequence 
are almost the same as CCSNe and PISNe, except for Fe; the reason is 
that in this model with the explosion energy 3 $\times$ $10^{52}$ erg, 
the innermost Fe--Si--S layer is not ejected because of a shallow 
mass cut caused by adjusting $M(^{56}{\rm{Ni}})$ in 
the ejecta to 0.07 $M_{\odot}$. 
The condensation time of each grain species is about 200 days earlier
than CCSNe, because the gas temperature quickly decreases due to the 
high expansion velocity and the low gas density in the ejecta. 
However, as can be seen from Fig. 15b, the average radius of each dust 
grains formed in the ejecta is almost the same as that in the ejecta of 
CCSNe, because the early condensation time compensates the decrease of gas 
density at the condensation time. 
 
With $M(^{56}{\rm Ni})=$ 0.7 $M_{\odot}$ and the higher gas temperature
at a time than H30A, as is shown in Fig. 16a, the condensation time of 
each grains species in the ejecta of H30B is about 150 days later than 
that of H30A.  
Therefore, the average radii of dust grains formed in the ejecta are a
factor 4--10 smaller than those of H30A (see Fig. 16b), depending on 
the concentration of key species at the formation site.  
In contrast to H30A, iron grain condenses in the innermost
Fe--Si--S layer, reflecting the deep mass cut at $M_r =$ 4.33 $M_{\odot}$.

From the results of calculations, we can conclude here that the explosion
energy as well as the amount of $^{56}$Ni in the ejecta strongly affects
the formation of dust grains in the ejecta. 
In the ejecta with a fixed amount of $^{56}$Ni, the higher explosion 
energy results in the early condensation time because the gas
temperature quickly decreases due to the low gas density. 
However, the average radius of dust grain is not heavily affected
because the early condensation time compensates the decrease of gas
density at the formation site. 
On the other hand, the mass of $^{56}$Ni in the ejecta strongly affects
the average radius of dust grains formed in the ejecta with the same
explosion energy.  
With increasing $M(^{56}{\rm Ni})$, the more delayed condensation time 
decreases the concentration of key species at the site of dust
formation, which results in the smaller size of dust grains. 
Finally it should be pointed out here that the amount of $^{56}$Ni
in the ejecta also affects the total mass of dust grains formed in the
ejecta as is depicted in Fig. 13a; a large amount of $^{56}$Ni in the
ejecta with a deep mass cut for a fixed progenitor mass result in 
the large mass in the ejecta as well as the large total mass of dust
grains, and vice versa.

\section{The effect of formation efficiency of CO and SiO molecules on 
dust formation}

In the calculations of dust formation, we assumed the complete formation 
of CO and SiO molecules. 
The formation of CO and SiO molecules prior to the dust formation was 
observed in SN 1987A (e.g., Bouchet \& Danziger 1993). 
Furthermore, the recent observations have suggested that formation of CO 
molecules is a common phenomenon in the ejecta of CCSNe (e.g., Gerardy et
al. 2002). 
Formation of CO and SiO molecules is usually considered to be complete 
and controls the chemistry in the astrophysical environments of interest 
to dust formation such as in the circumstellar envelopes of AGB star. 
However, in the ejecta of supernovae, these molecules are destroyed 
by the collisions with high energy electrons as well as with ionized 
inert gaseous atoms produced by the decay of radio active elements, and 
the abundance of these molecules is determined by the balance between 
formation and destruction (Liu \& Dalgarno 1994, 1996). 
Thus, in the ejecta of supernovae, for an example, it is expected that 
free carbon atoms are available for formation of carbon--bearing dust 
grains even in the ejecta of C/O $<$ 1. 
This aspect has been extensively pursued by Clayton, Liu, \& Dalgarno 
(1999) and Clayton, Deneault, \& Meyer (2001) 
to investigate the origin of presolar grains identified as SUNOCONs. 
The abundances of CO and SiO molecules depend on the abundance of free
electrons as well as ionized inert gas. Thus, in this Appendix, 
to simplify the calculations, we investigate the effect of the
formation efficiency of CO and SiO molecules on the dust formation, 
by referring to the formation efficiency of CO and SiO molecules observed in
SN 1987A as a template and introducing the parameters $f_{\rm C}$ and
$f_{\rm Si}$ which represent the mass fractions of C and Si atoms not
locked into CO and SiO molecules, respectively. 

In the ejecta of SN 1987A, the masses of CO and SiO molecules have been 
evaluated to be 3--6 $\times 10^{-3}$ $M_{\odot}$ and 0.4--1 $\times
10^{-3}$ $M_{\odot}$, respectively, during 300 to 600 days after 
the explosion (Liu \& Dalgarno 1994, 1995), which correspond to 
$f_{\rm C} \simeq 0.99$ and $f_{\rm Si} \simeq 0.9$ in the oxygen--rich 
core. The mass fractions $f_{\rm C} = 0.99$ and 
$f_{\rm Si} = 0.9$ being fixed, respectively, Figs. 17a and 17b show 
the mass of newly formed grain species versus $f_{\rm Si}$ and $f_{\rm
C}$ at the location of $M_r =$ 3.5 $M_{\odot}$ in the O--Mg--Si 
layer of C20.  In the calculations, we consider SiO molecule is the key 
species for formation of Mg--silicate grains. Of course, formation of 
Al$_2$O$_3$ grain is not affected by the formation efficiency of these 
molecules. The mass of C and Si grains increases with increasing $f$. 
As can be seen from Fig. 17a, except for the formation of carbon grains,
formation efficiency of CO molecules does not influence the formation
of other grain species. Also the formation of carbon grains does not
affect the total mass of dust grains at this location since O and Mg 
atoms are much more abundant than C atoms (see Fig. 1a). The average 
size of carbon grains increases with increasing $f_{\rm C}$ but is 
limited to less than 0.04 $\mu$m even for $f_{\rm C} = 1$. 
On the other hand, the formation efficiency of SiO molecules strongly 
affects the mass abundance of Si-- and/or Mg--bearing grains;  
with increasing $f_{\rm Si}$, the 
decrease of SiO molecules not only results in the formation of Si grains 
but also increases the mass of MgO grains with the reduction of mass of 
Mg$_2$SiO$_4$ grains, which is prominent for $f_{\rm Si} \ge 0.1$. 
The average radii of Si and MgO grains increase and reach to 
$\sim 0.1$ $\mu$m with increaing $f_{\rm Si}$, otherwise the average 
radius of Mg$_2$SiO$_4$ grain decreases. At this location, formation 
efficiency of CO and SiO molecule does not affect the total mass of 
dust grains because the abundance of C atom is two order of magnitude 
lower than that of O atom. However, if $f_{\rm C} = 0.99$ in the outer 
region of O--Mg--Si layer (4 $M_{\odot} \le M_r \le 5 ~M_{\odot}$ for 
C20), much more carbon dust grains could condense and the average radius 
reach to 0.4 $\mu$m $\sim 1 ~\mu$m, depending on the abundance of carbon 
atom at the formation site. 

Finally it should be addressed here that we cannot realize the formation
of SiC grains with radius comparable to presolar SiC grains 
identified as SUNOCONs by changing the parameters $f_{\rm C}$ and
$f_{\rm Si}$. Formation of SiC grains competes with the formation of carbon 
and silicon grains; carbon atoms available for formation of SiC grains are
locked into carbon grains prior to formation of SiC grains, which is
true for silicon atoms. This implies that the amount and size are
very small even if SiC grains condense. Therefore, as proposed by 
Deneault, Clayton, \& Heger (2003), a special condition and/or chemical 
passways might have to be considered in order to realize the formation
of SiC grains identified as SUNOCONs. 
 
Anyway, the result of calculation shows that the reduction of formation
efficiency of CO and SiO moclules leads to the formation of C and Si grains. 
Furthermore, the formation efficiency of SiO molecules affects the
abundance and the size of Si-- and/or Mg--bearing grains.  
However we must keep in mind that in the calculations the stability of C 
and Si grains is not taken into account; in the environment with a lot 
of free oxygen atoms, C and Si grains could be easily oxidized if their 
sizes would be small, though Clayton, Deneault, \& Meyer (2001) 
have claimed that carbon grain is stable against the oxidation in 
the oxygen--rich environment. Oxidation of Si grains produces SiO$_2$ 
grains and/or evaporates SiO molecules available for formation of 
Mg--silicate grains, and it may be possible that another chemical
passway leads to formation of Mg--silicate grains without SiO molecules, 
which results in the more oxide grains even if the formation efficiency 
is very low. These aspects should be pursued to clarify how much 
as well as what kind of oxide grains really forms in 
the ejecta of supernovae.

\clearpage

\begin{figure}
\epsscale{0.8}
\plotone{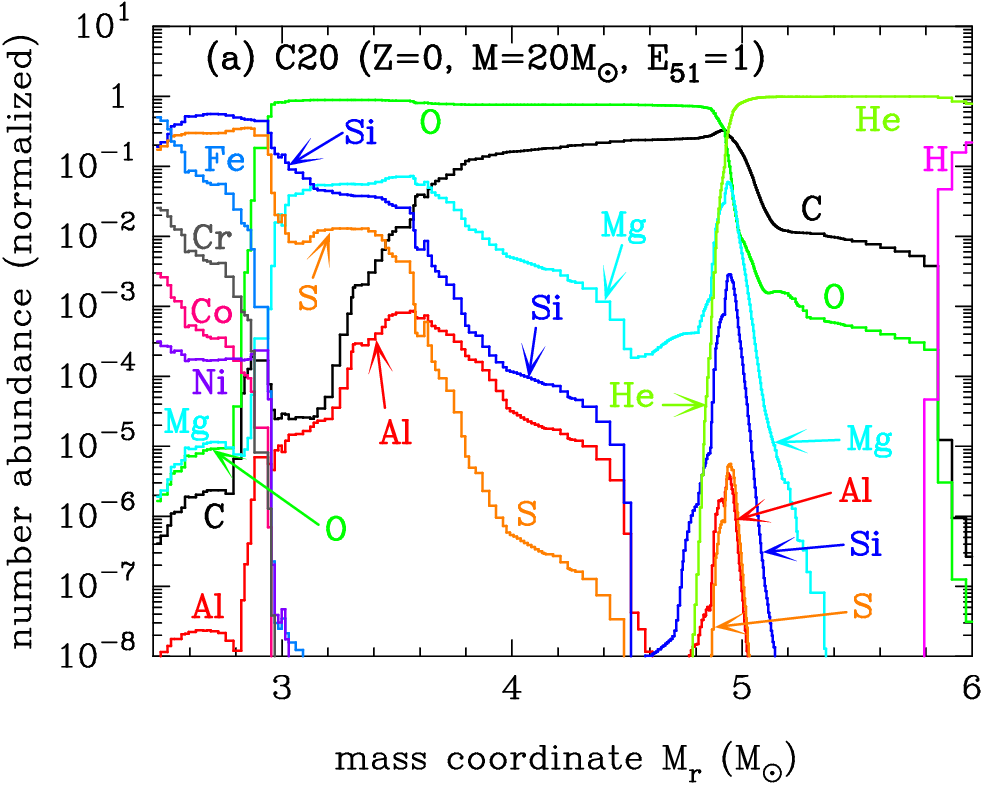}
\plotone{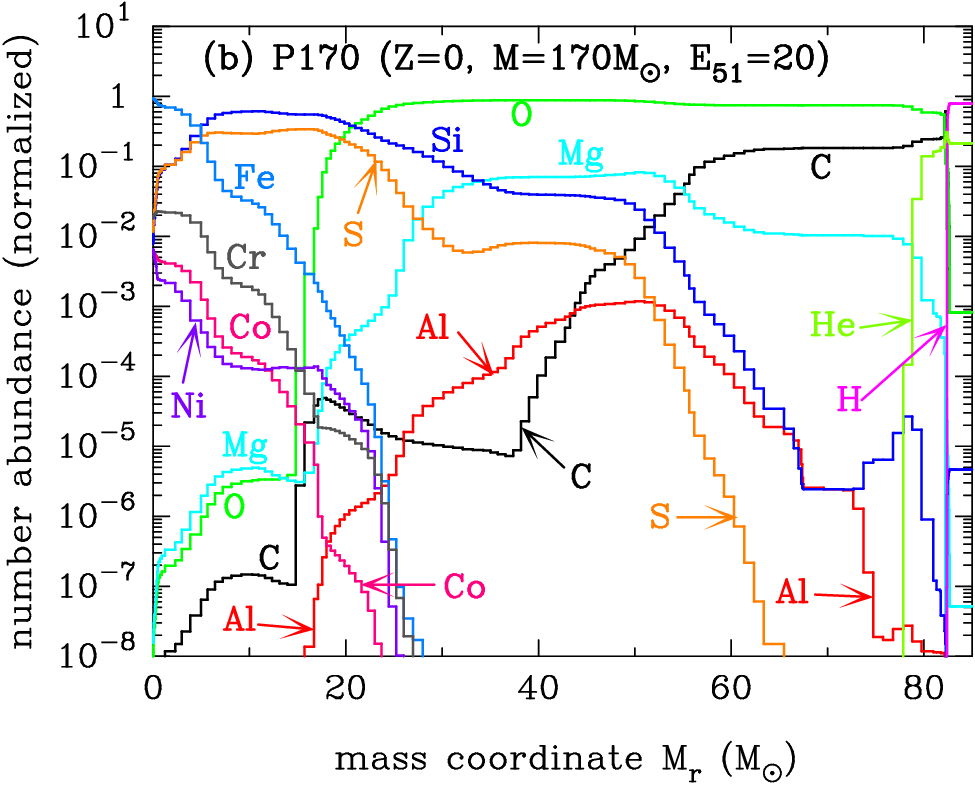}
\caption{The number abundance of elements relevant to dust formation
 within the unmixed He--core at 600 days after the explosion, taking 
 into account the decay of radio active elements; (a) for C20 and (b) for 
 P170.
[{\it See the electric
 edition of the Journal for a color version of this figure.}]
\label{fig1}}
\end{figure}

\clearpage

\begin{figure}
\plotone{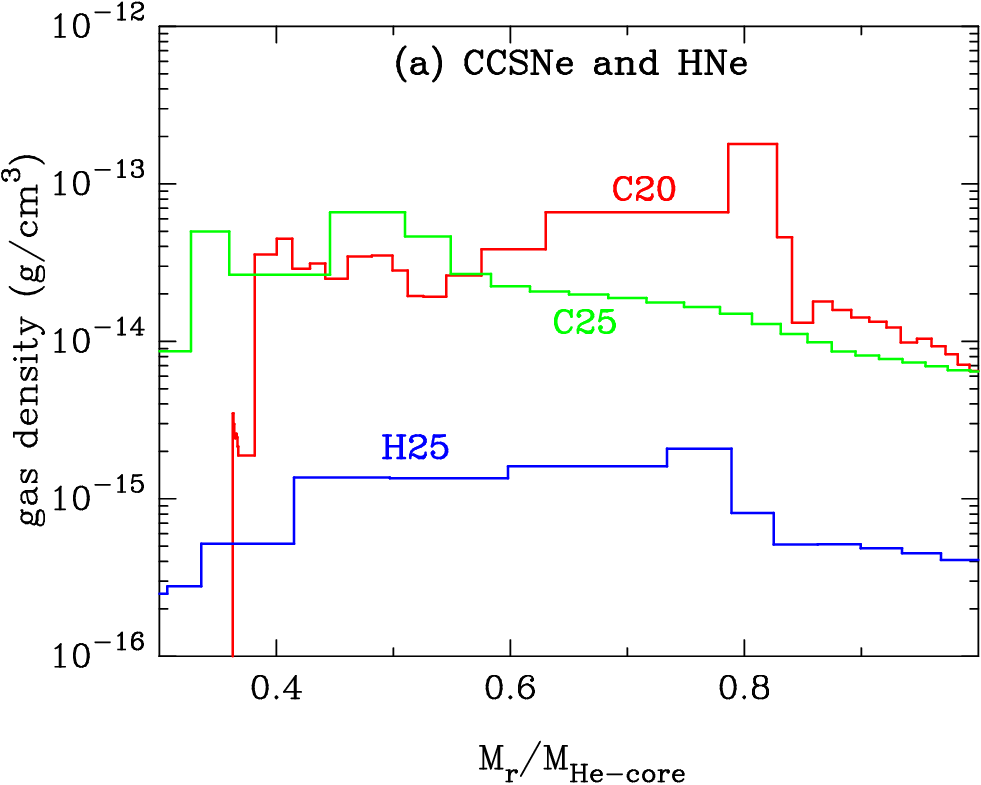}
\plotone{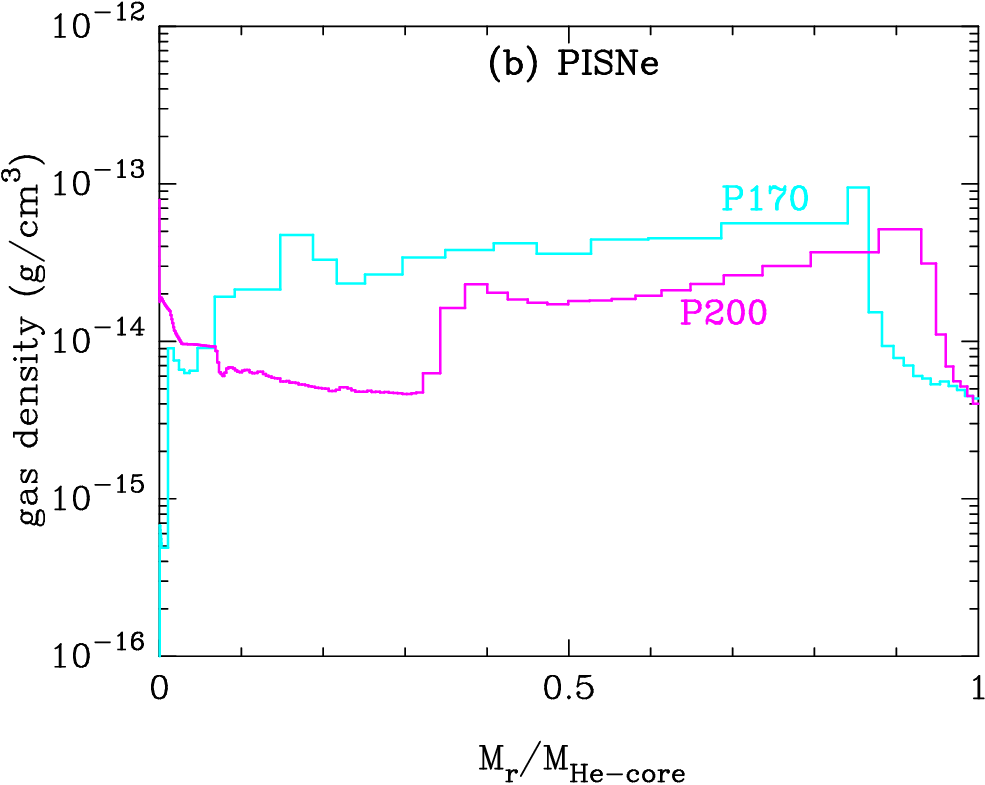}
\caption{The distribution of gas density within the He--core at 600 days
 after the explosion; (a) for C20, C25 and H25 and (b) for P170 and P200. 
 Note that the mass coordinate is normalized to the He--core mass.\label{fig2}}
\end{figure}

\clearpage

\begin{figure}
\plotone{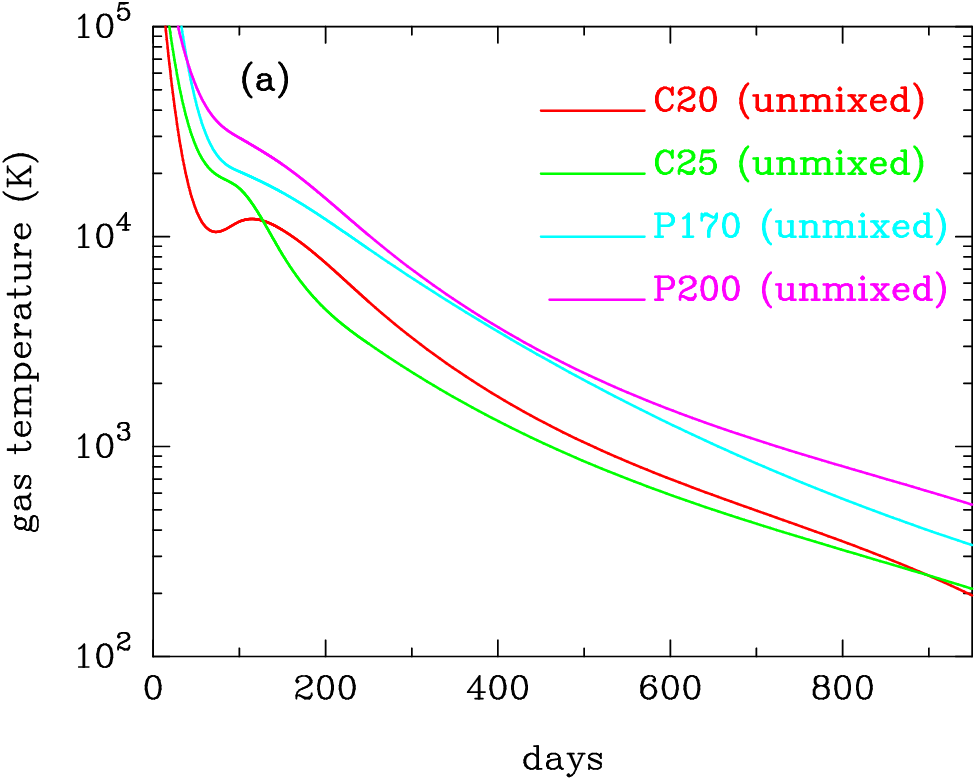}
\plotone{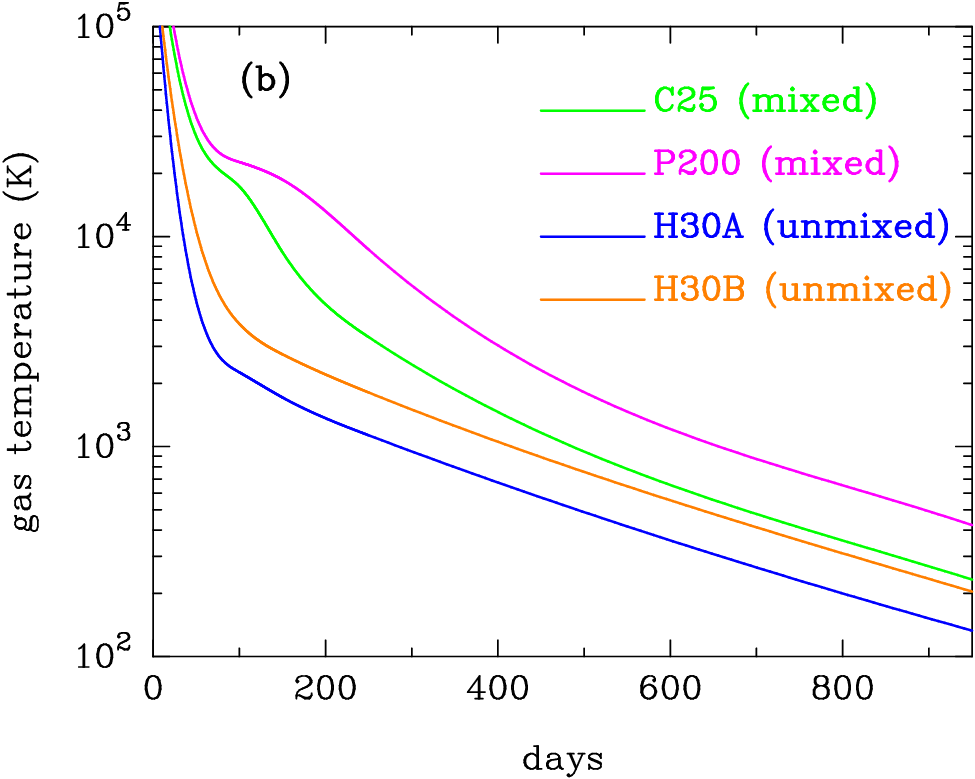}
\caption{The time evolution of gas temperature at a location of the 
 oxygen--rich layer with almost the same elemental composition in the
 ejecta of CCSNe, PISNe and HNe; (a) at $M_r=$ 3.7 $M_{\odot}$ for C20, 
 at $M_r=$ 4.6 $M_{\odot}$ for C25, at $M_r=$ 56.5 $M_{\odot}$ for P170, 
 and at $M_r=$ 64.9 $M_{\odot}$ for P200 in the unmixed ejecta, 
 (b) at $M_r=$ 4.6 $M_{\odot}$ for C25 and at $M_r=$ 64.9 $M_{\odot}$
 for P200 in the mixed ejecta, and at $M_r=$ 6.8 $M_{\odot}$ for H30A 
 and at $M_r=$ 6.8 $M_{\odot}$ for H30B in the unmixed ejecta.\label{fig3}}
\end{figure}

\clearpage

\begin{figure}
\plotone{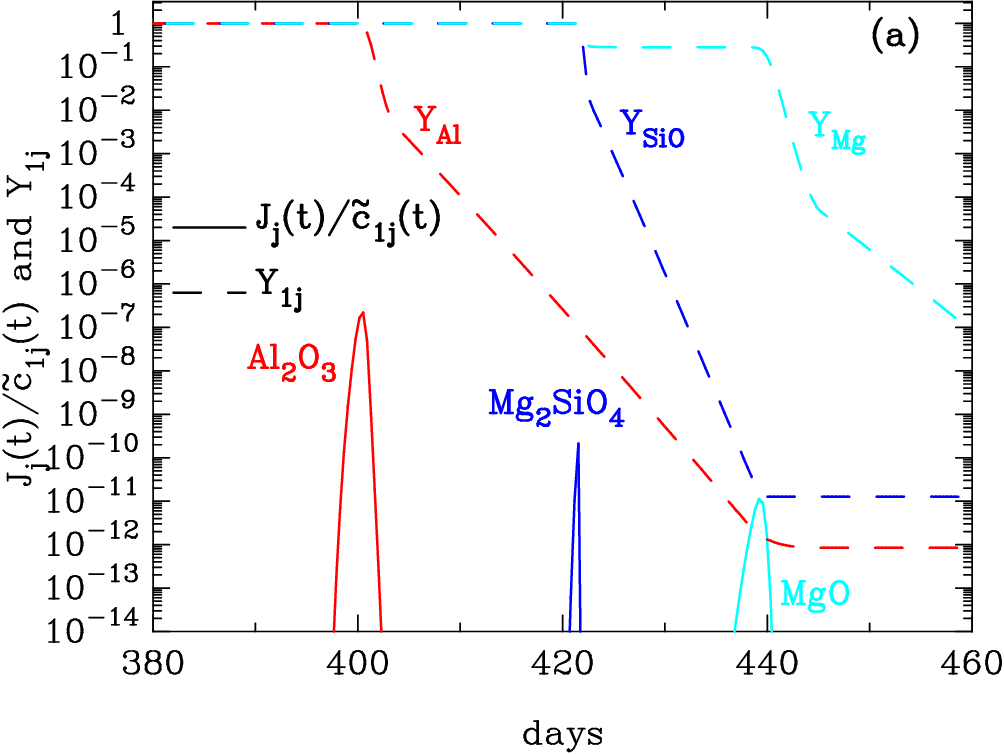}
\plotone{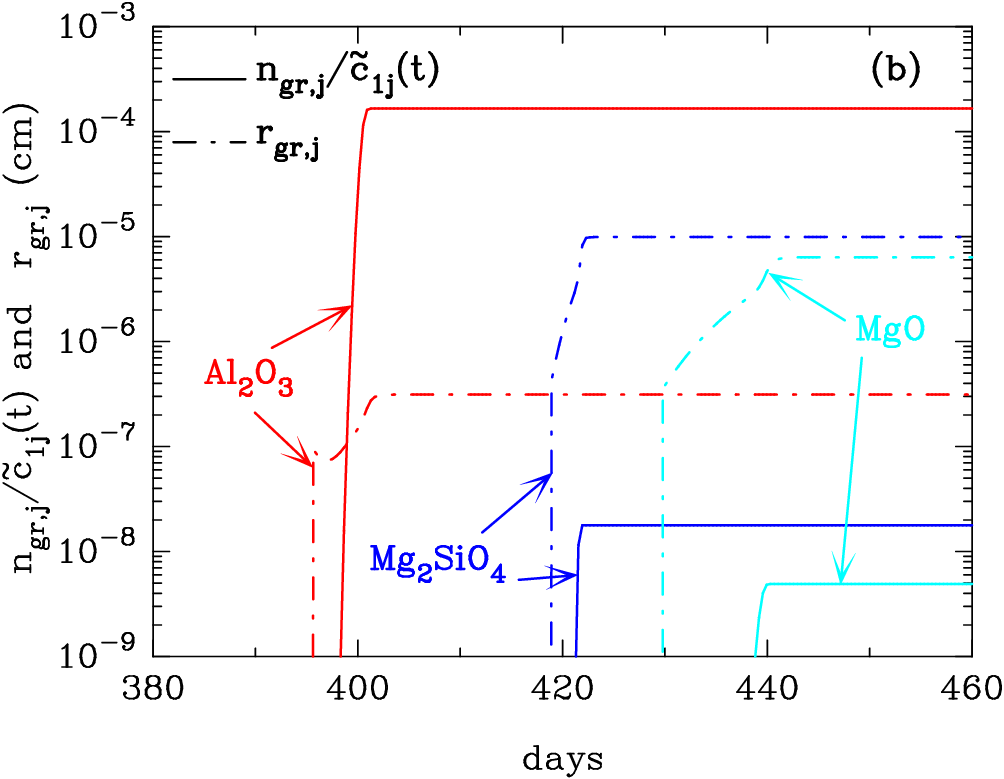}
\caption{The behaviors of nucleation and growth of each grain species
 condensed at $M_r=$ 3.5 $ M_{\odot}$ in the O--Mg--Si layer of the
 unmixed ejecta for C20; (a) the time evolution of the nucleation rate 
 $J_j(t)$ normalized to  $\tilde{c_{1j}}(t)$ (solid line) and the
 depletion of key species $Y_{1j}$ (dashed line), 
 (b) the time evolution of the number density $n_{{\rm gr},j}$
 normalized to $\tilde{c_{1j}}(t)$ (solid line) and the average grain 
 radius $r_{{\rm gr},j}$ (dot--dashed line).
 The number abundances of dust forming elements relative to oxygen 
 at this location are Si/O $=2.97 \times 10^{-2}$, Mg/O $=8.25 \times 
 10^{-2}$ and Al/O $=9.38 \times 10^{-4}$.\label{fig4}}
\end{figure}

\clearpage 

\begin{figure}
\plotone{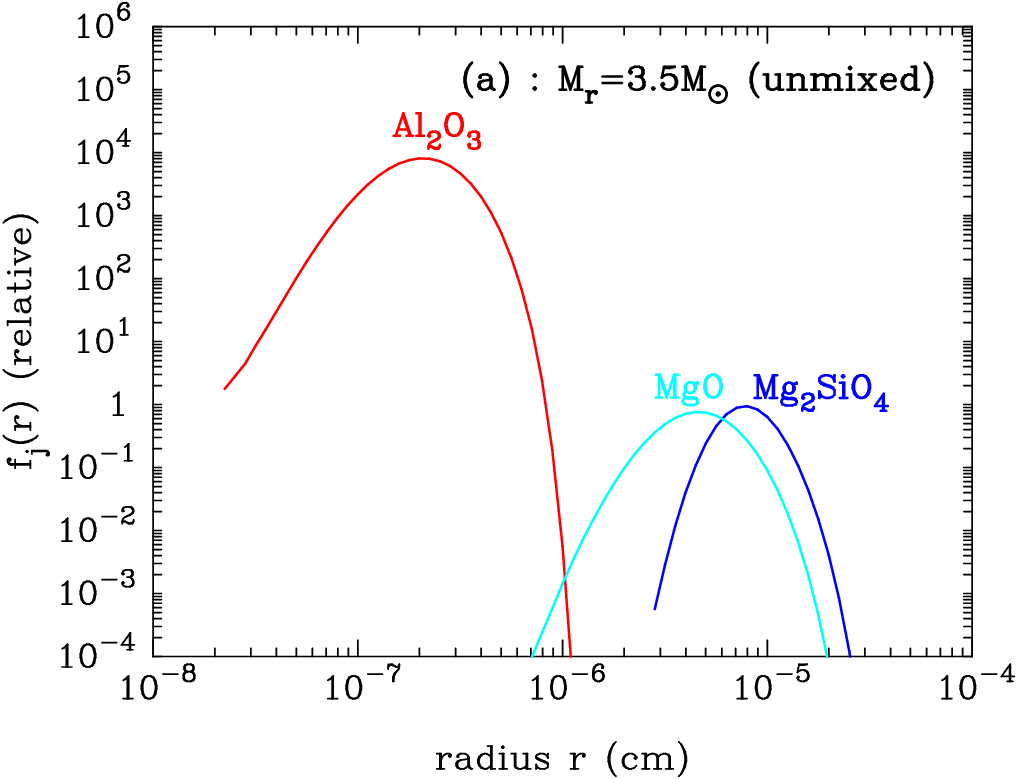}
\plotone{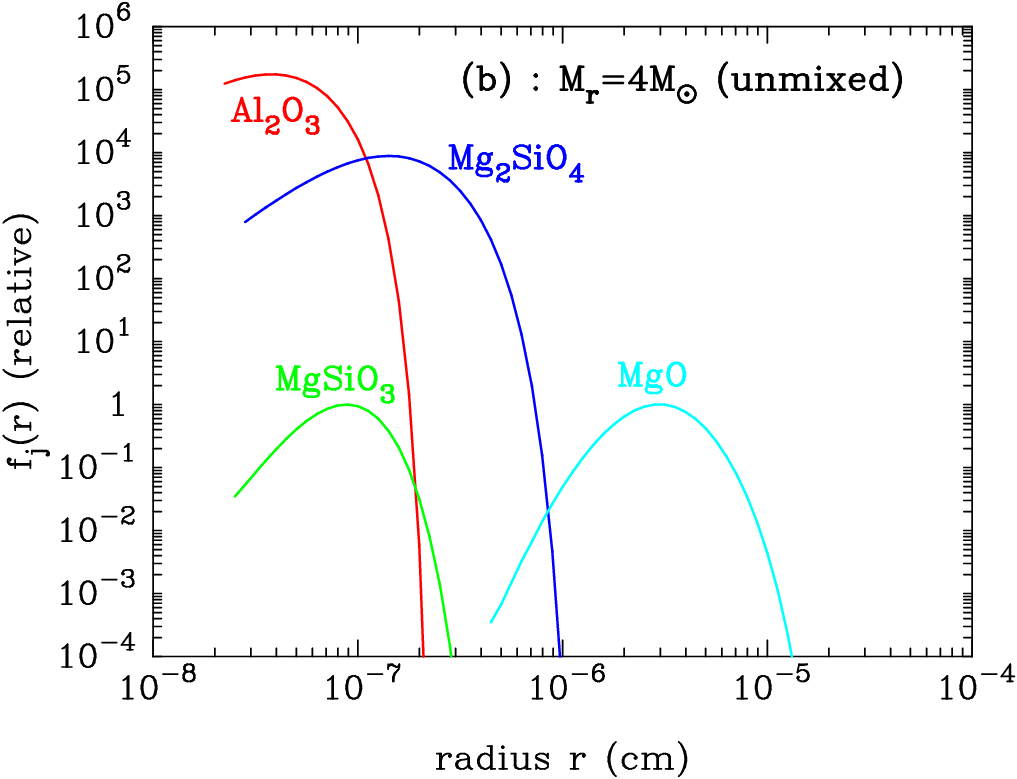}
\caption{The size distribution function of each grain species formed at 
 a location in the O--Mg--Si layer of the unmixed ejecta for C20; (a) at 
 $M_r=$ 3.5 $M_{\odot}$ and (b) at $M_r=$ 4 $M_{\odot}$ with Si/O 
 $ =1.50 \times  10^{-4}$, Mg/O $ =6.40 \times 10^{-3}$, and Al/O $ = 
 3.95 \times 10^{-5}$. 
\label{fig5}}
\end{figure}

\clearpage 

\begin{figure}
\plotone{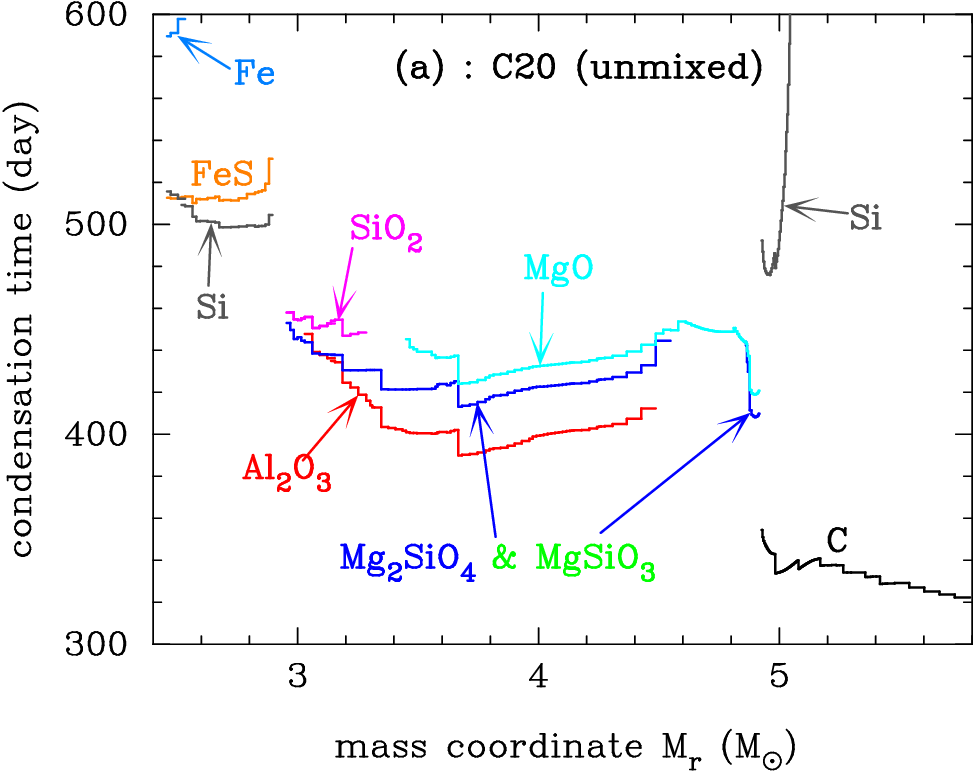}
\plotone{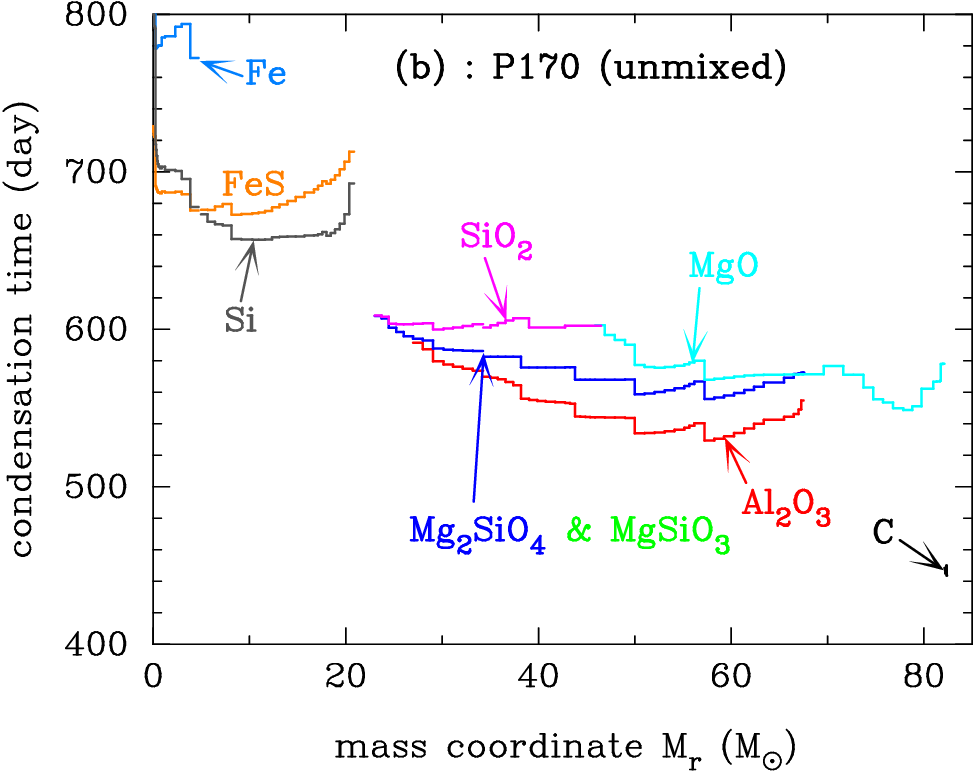}
\caption{The condensation times of dust grains formed in the unmixed 
 ejecta; (a) for C20 and (b) for P170.
[{\it See the electric
 edition of the Journal for a color version of this figure.}]
\label{fig6}}
\end{figure}

\clearpage 

\begin{figure}
\plotone{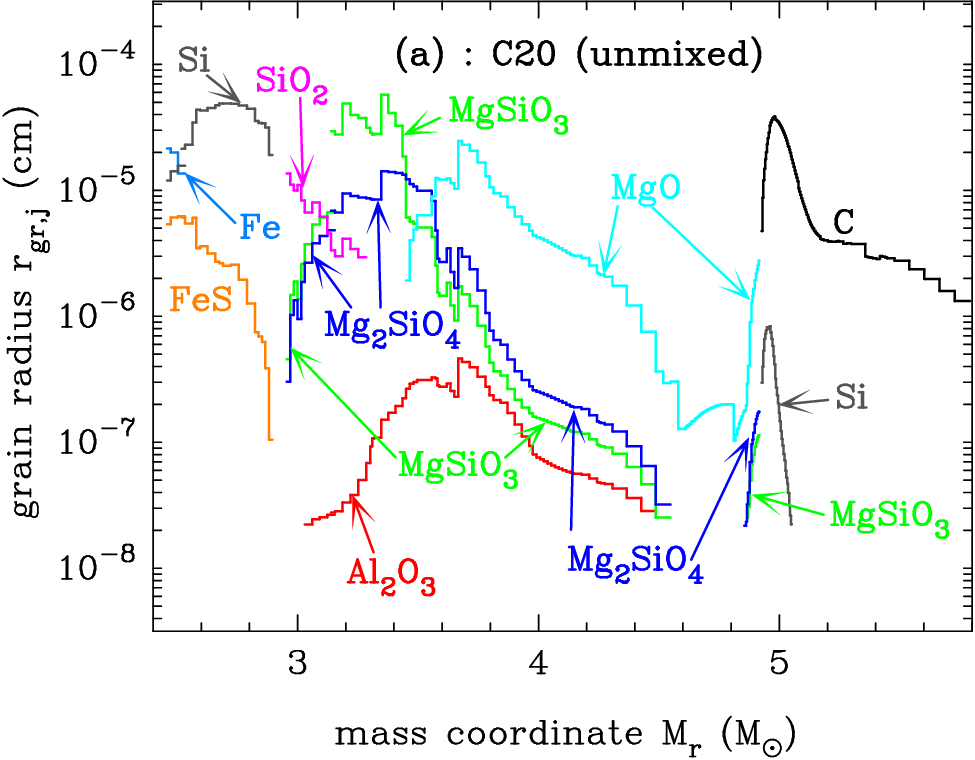}
\plotone{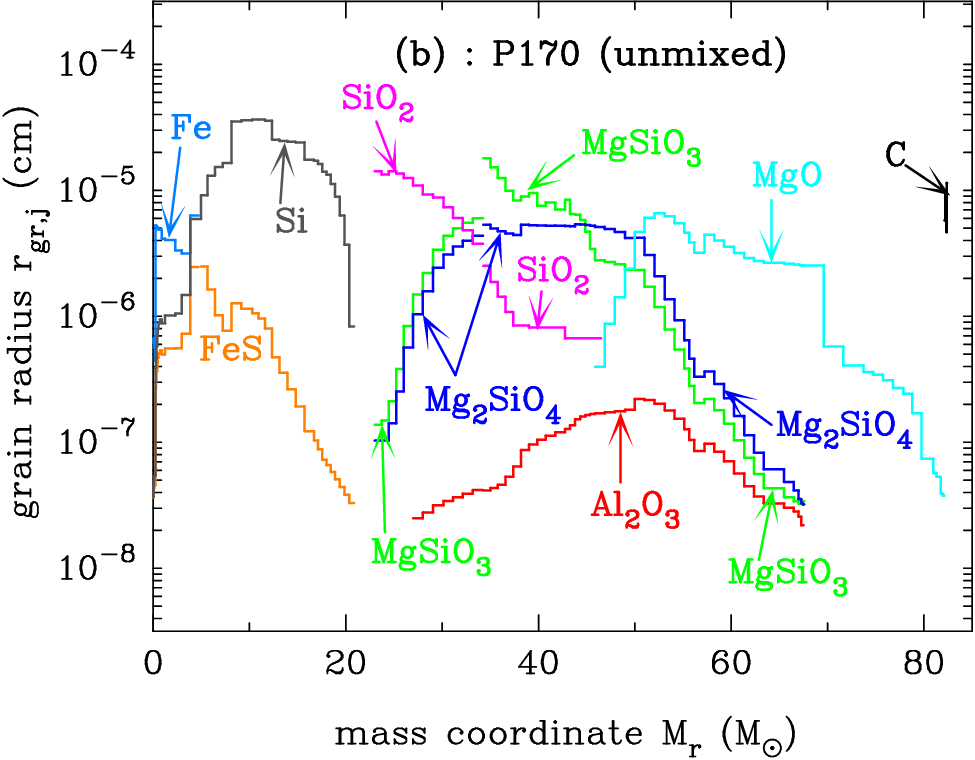}
\caption{The average radii of dust grains formed in the unmixed ejecta;
 (a) for C20 and (b) for P170. 
[{\it See the electric
 edition of the Journal for a color version of this figure.}]
\label{fig7}}
\end{figure}

\clearpage 

\begin{figure}
\plotone{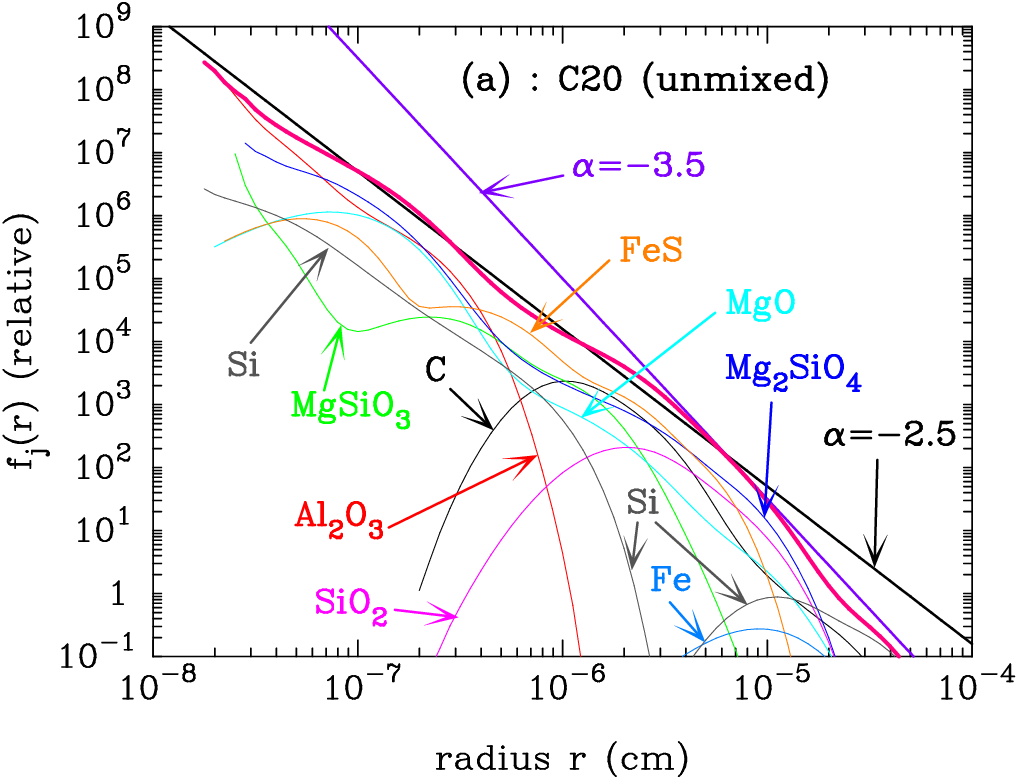}
\plotone{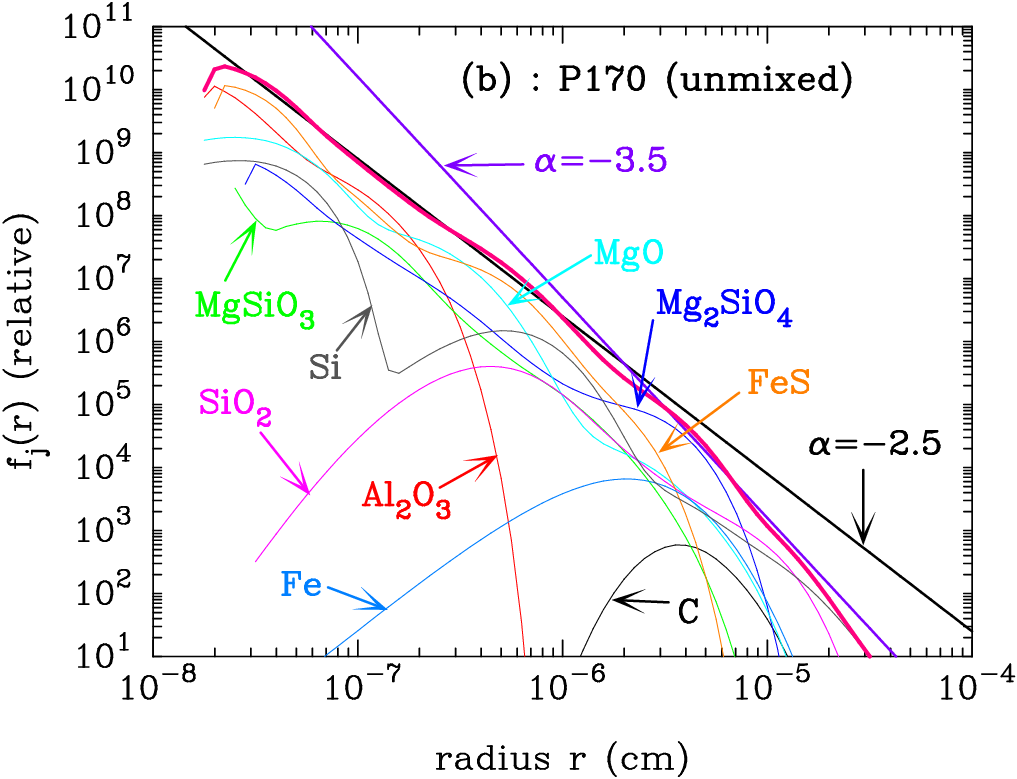}
\caption{The size distribution function of each grain species 
 summed up over the
 formation region in the unmixed ejecta. The thick curve represents 
 the size distribution function summed up over all grain species and 
 the straight lines indicate power--law formulae with the index of
 $\alpha = -2.5$ and $\alpha = -3.5$; (a) for C20 and (b) for 
 P170.
[{\it See the electric
 edition of the Journal for a color version of this figure.}]
\label{fig8}}
\end{figure}

\clearpage 

\begin{figure}
\plotone{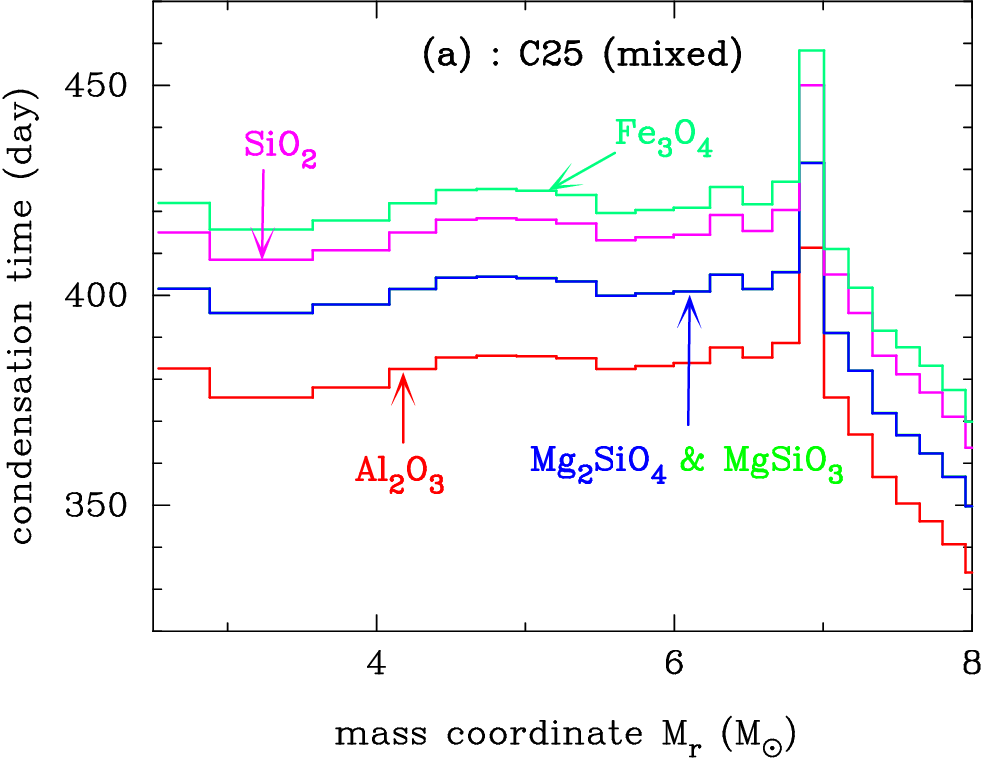}
\plotone{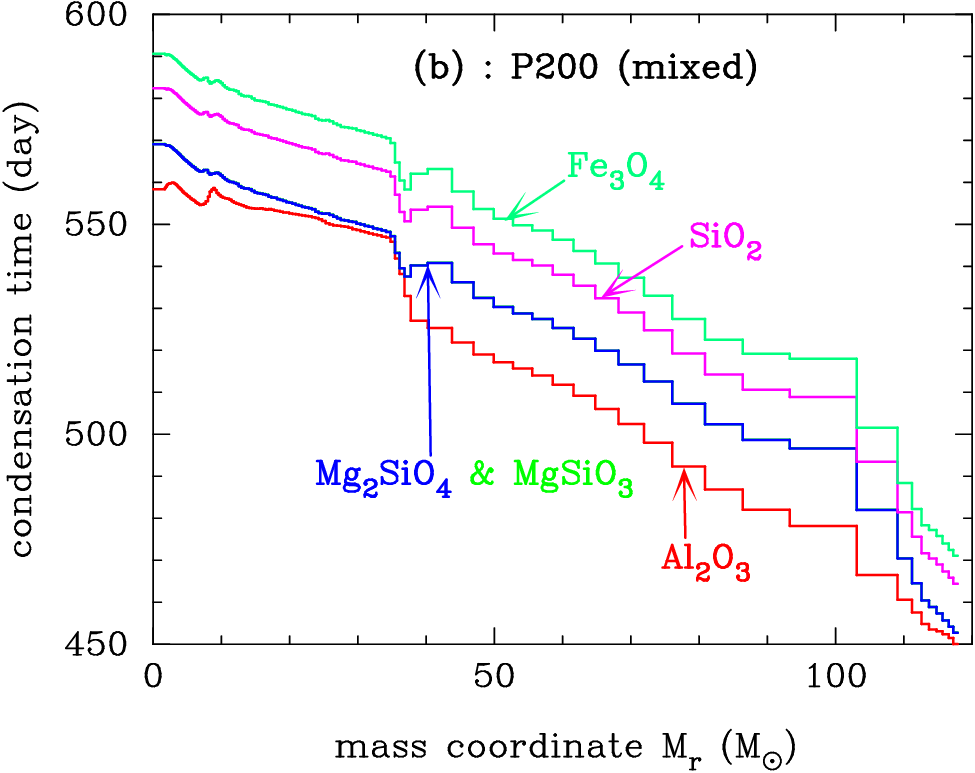}
\caption{The condensation times of dust grains formed in the mixed
 ejecta; (a) for C25 and (b) for P200.
[{\it See the electric
 edition of the Journal for a color version of this figure.}]
\label{fig9}}
\end{figure}

\clearpage 

\begin{figure}
\plotone{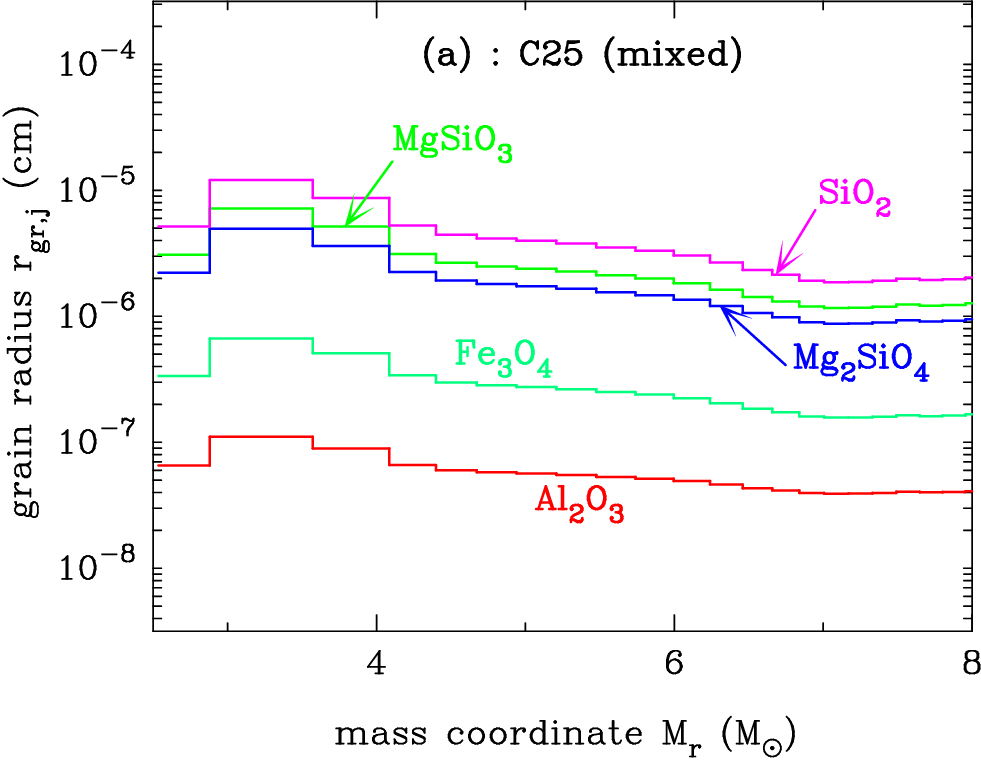}
\plotone{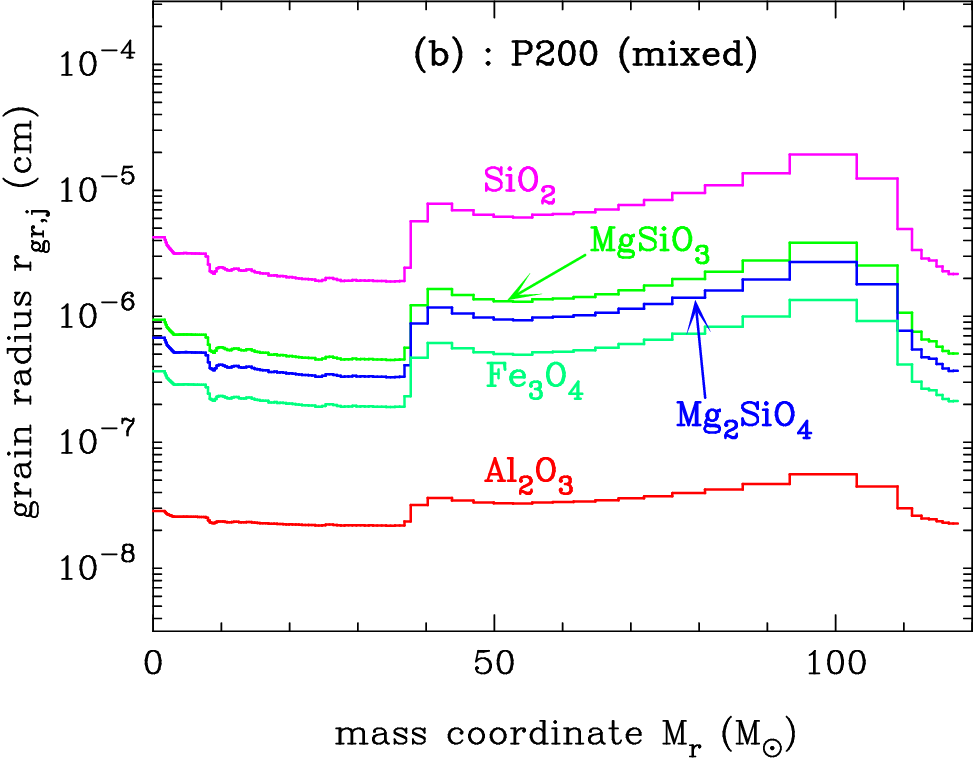}
\caption{The average radii of dust grains formed in the mixed ejecta;
 (a) for C25 and (b) for  P200.
[{\it See the electric
 edition of the Journal for a color version of this figure.}]
\label{fig10}}
\end{figure}

\clearpage 

\begin{figure}
\plotone{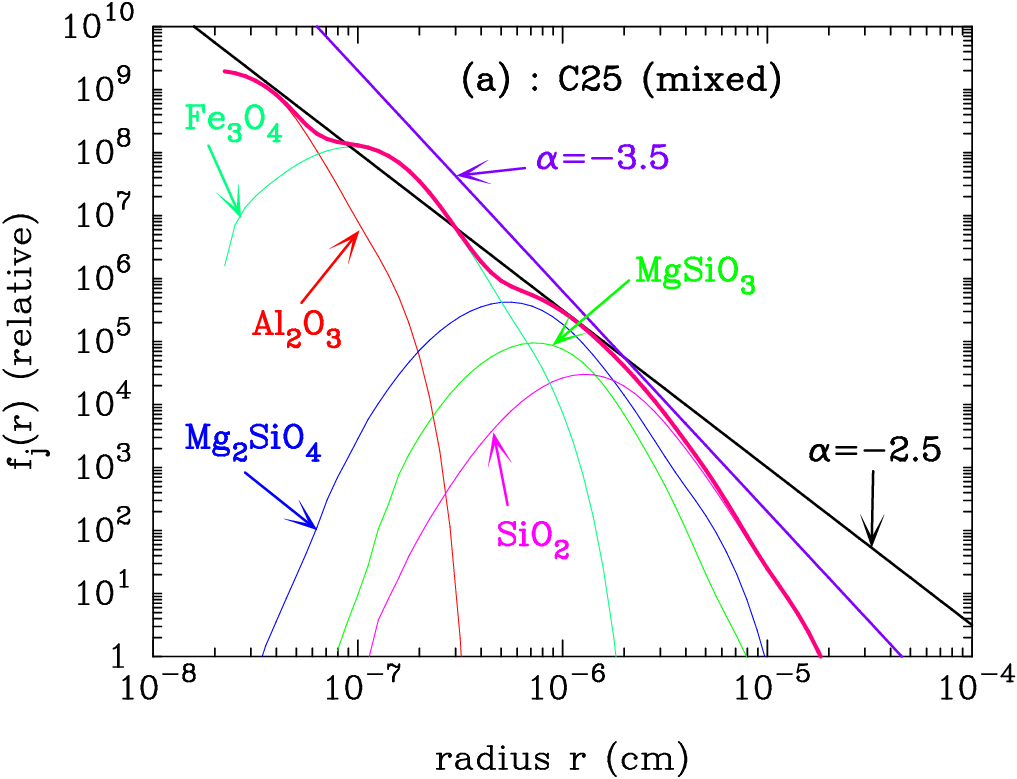}
\plotone{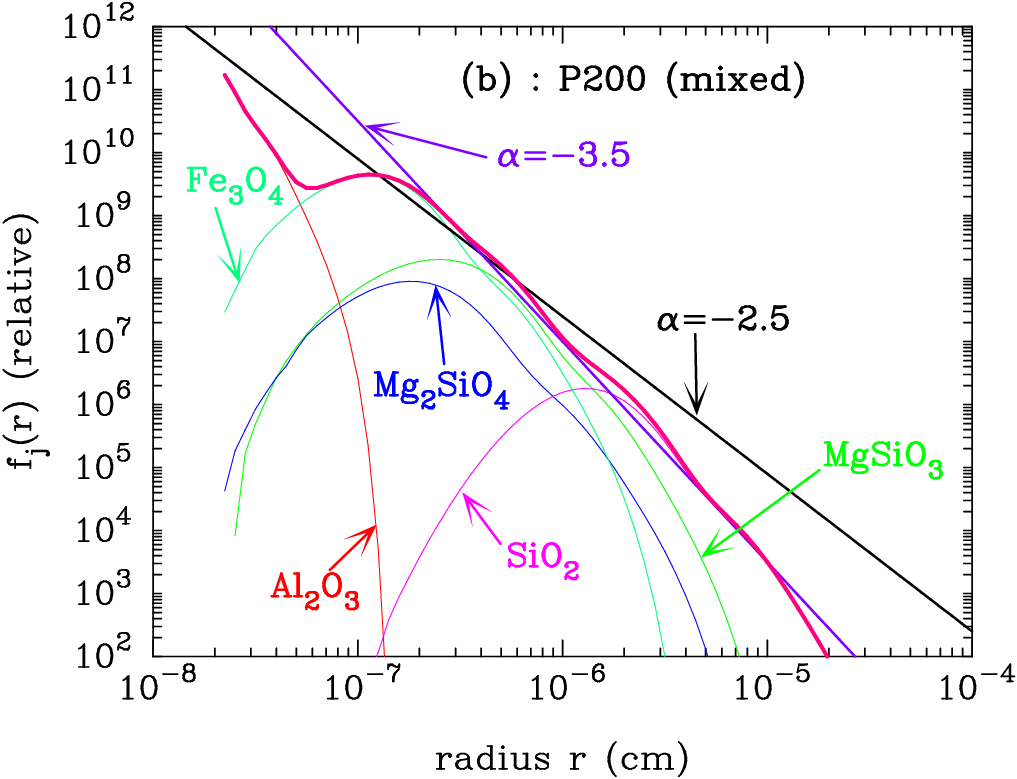}
\caption{The same as Fig. 8 but in the uniformly mixed ejecta; (a) for
 C25 and (b) for P200. 
[{\it See the electric
 edition of the Journal for a color version of this figure.}]
\label{fig11}}
\end{figure}

\clearpage 

\begin{figure}
\plotone{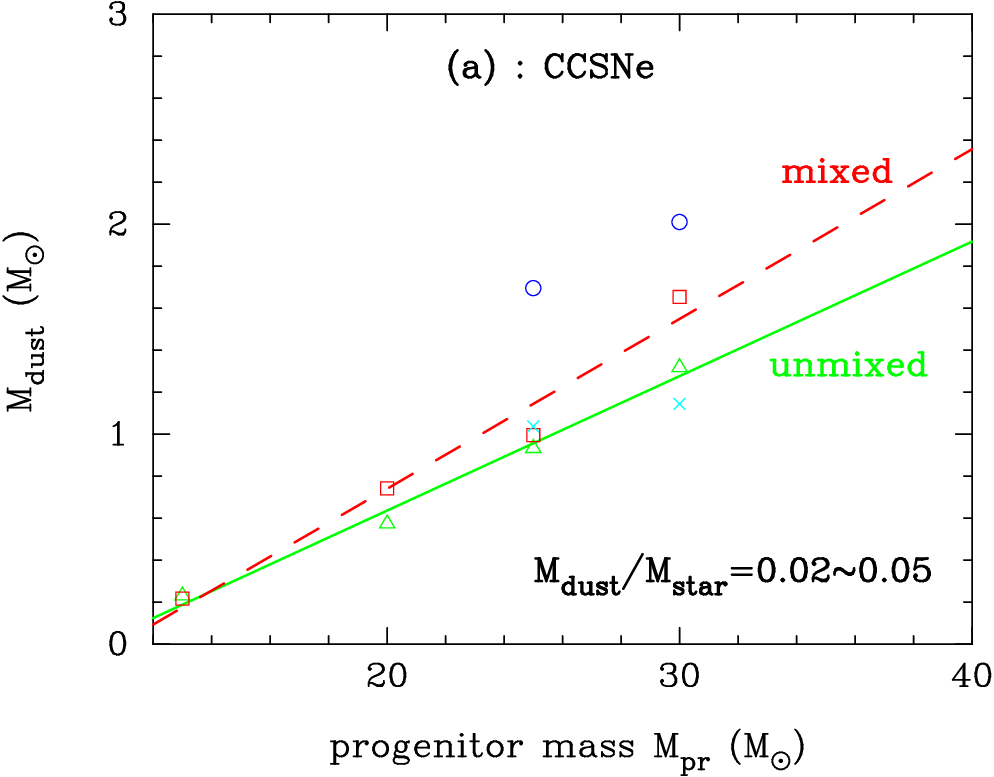}
\plotone{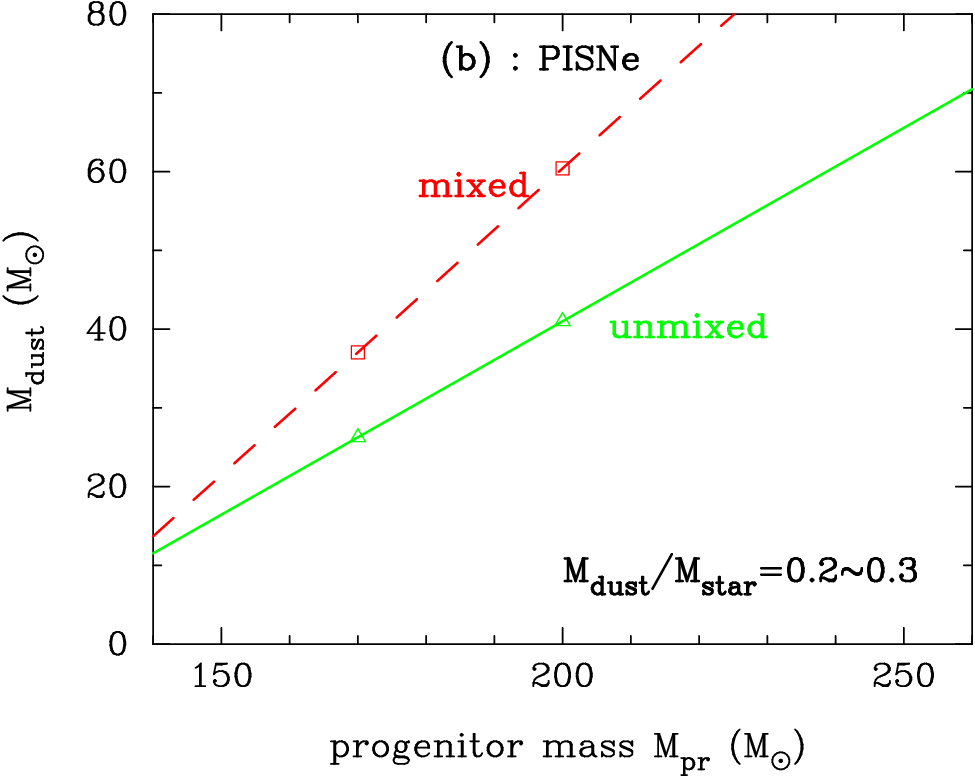}
\caption{The total mass of dust grains formed in the unmixed (open
 triangles) and mixed (open squares) ejecta of CCSNe and PISNe versus the 
 progenitor mass $M_{\rm pr}$; 
 (a) for CCSNe and (b) for PISNe. The straight lines indicate the least 
 square's fits to the calculated mass in the unmixed ejecta (solid line) 
 and the mixed ejecta (dashed line) of CCSNe (Fig. 12a), and
 connect the calculated mass for PISNe (Fig. 12b). 
 Also, in Fig. 12a the total mass of dust grains produced in the unmixed
 ejecta of hypernovae is plotted by the crosses (H25A and H30A) and the 
 open circles (H25B and H30B).\label{fig12}}
\end{figure}

\clearpage 

\begin{figure}
\plotone{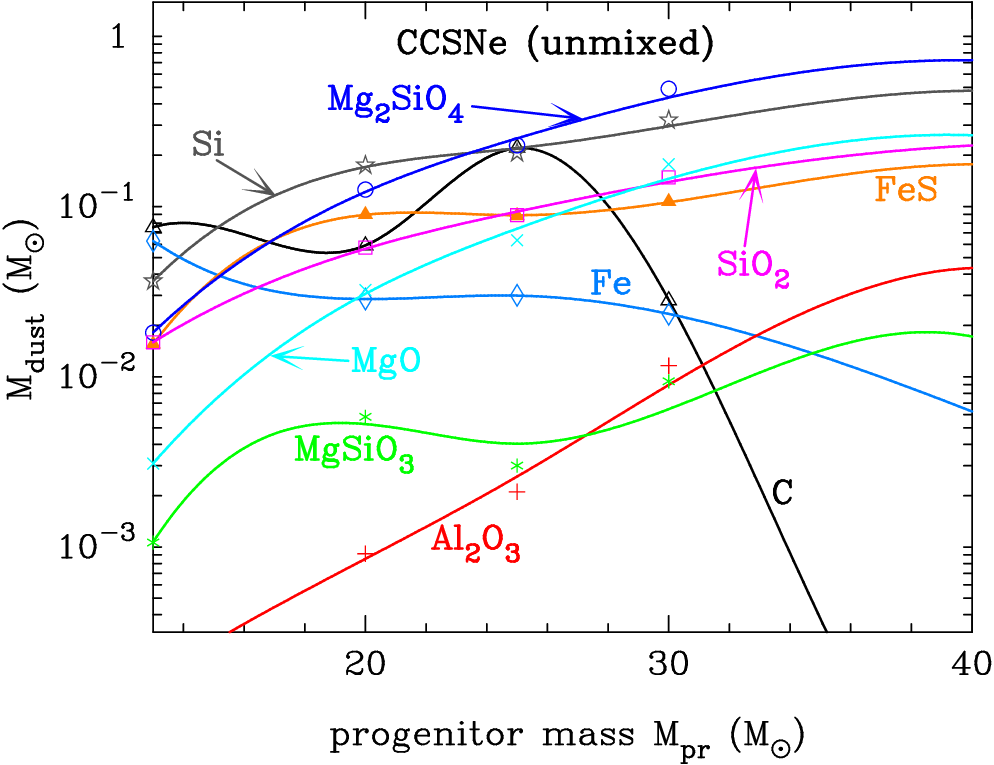}
\plotone{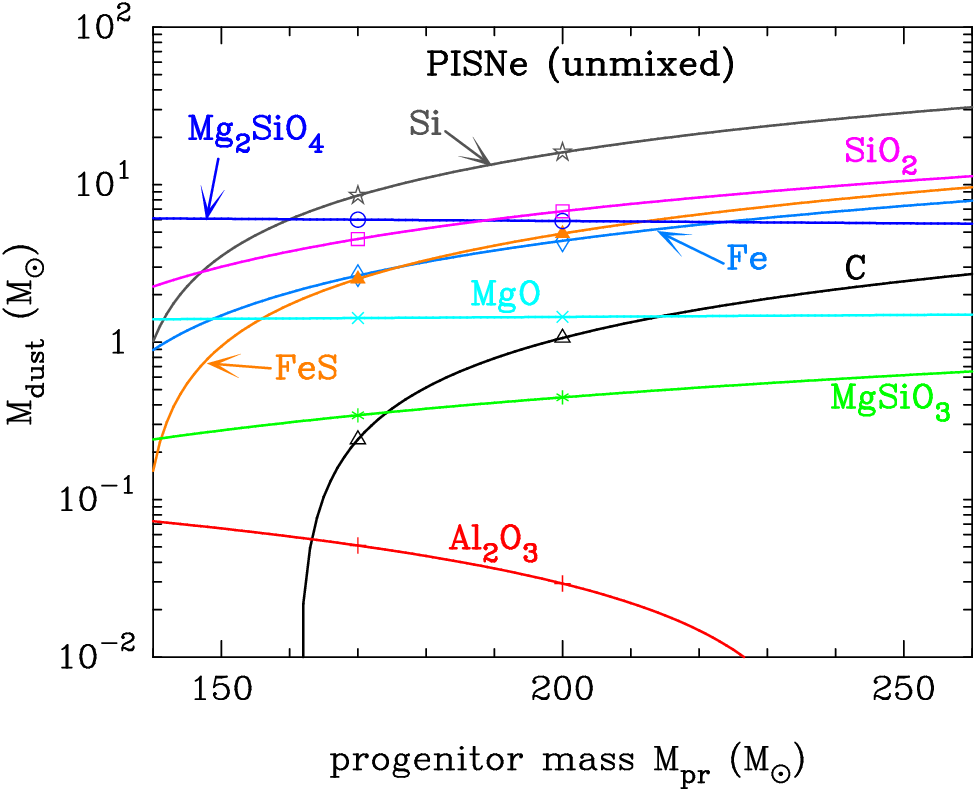}
\caption{The mass yield of each dust grain formed in the unmixed ejecta; 
 (a) for CCSNe and (b) for PISNe. The smooth curves are the least
 squares' spline fits to the calculated yields for CCSNe (Fig. 13a), and 
 are the straight lines connecting the calculated yields in linear scale for 
 PISNe (Fig. 13b).
[{\it See the electric
 edition of the Journal for a color version of this figure.}]
\label{fig13}}
\end{figure}

\clearpage 

\begin{figure}
\plotone{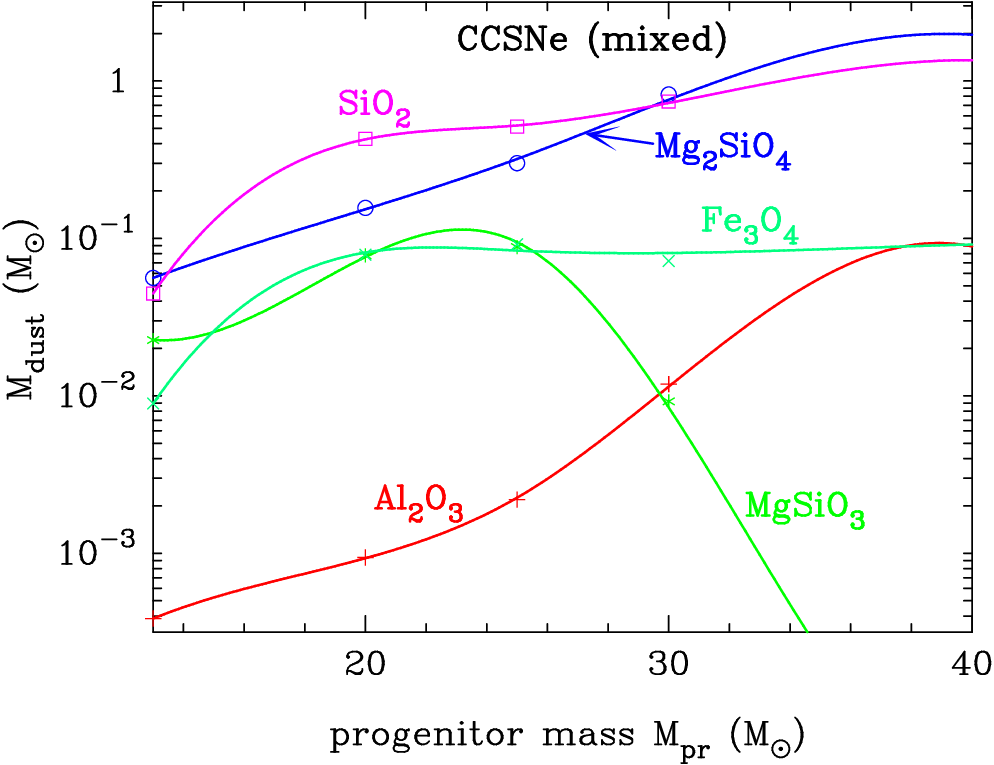}
\plotone{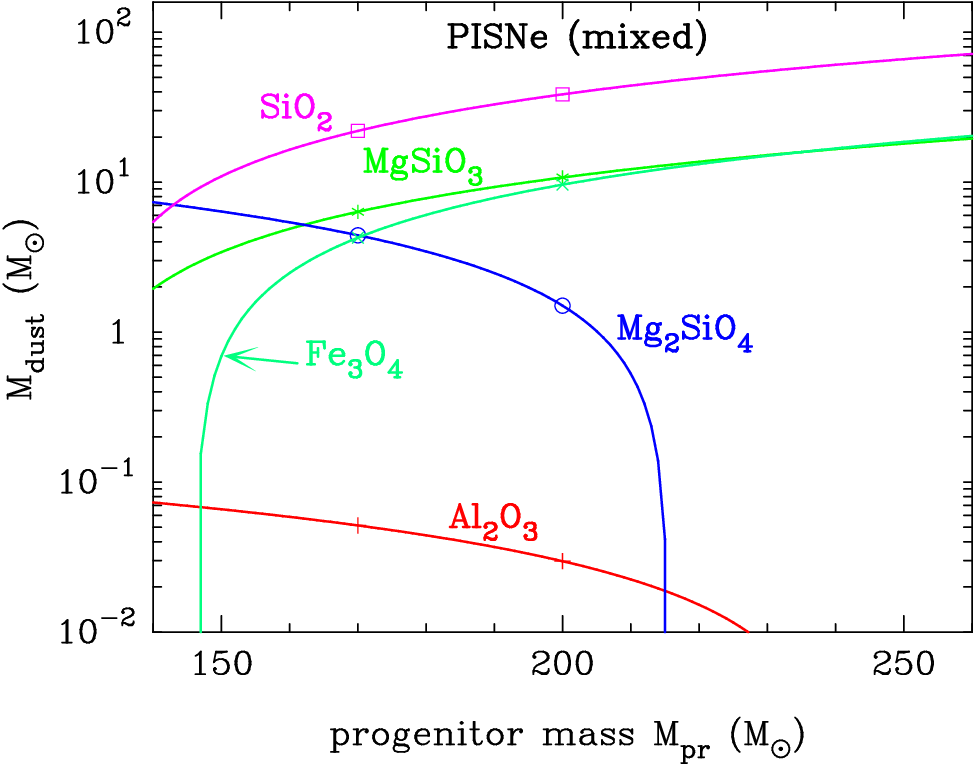}
\caption{The same as Fig. 13, but in the mixed ejecta; (a) for CCSNe, 
 and (b) for PISNe.
[{\it See the electric
 edition of the Journal for a color version of this figure.}]
\label{fig14}}
\end{figure}

\clearpage 

\begin{figure}
\plotone{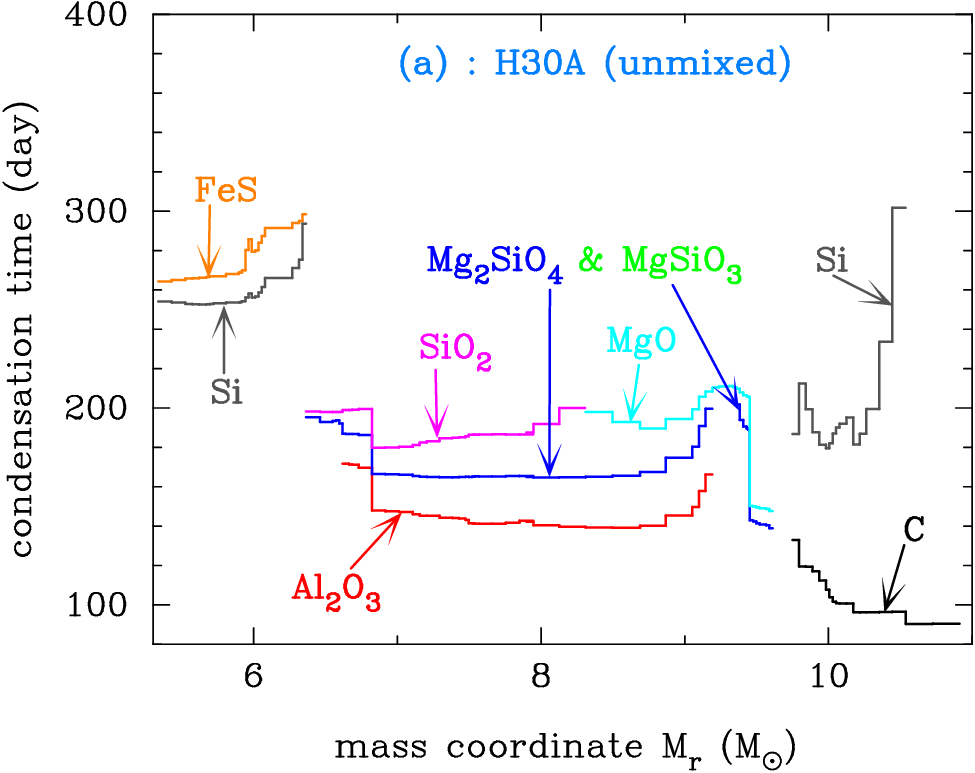}
\plotone{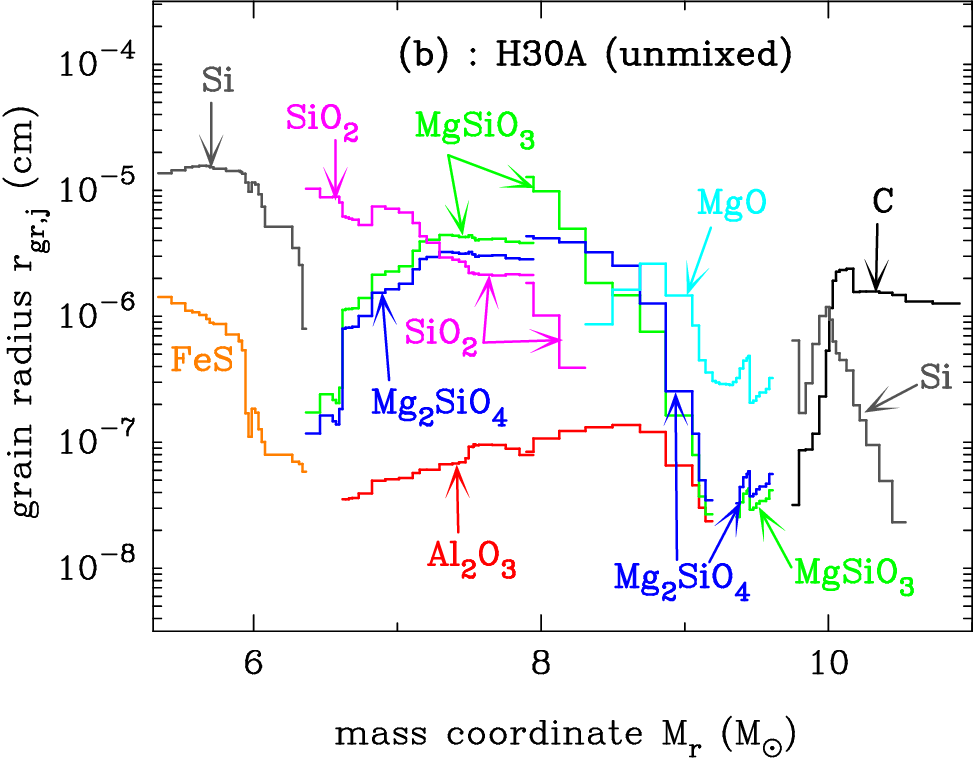}
\caption{The condensation times and the average radii of dust grains 
 formed in the unmixed ejecta of H30A; (a) the condensation times and 
 (b) the average radii. 
[{\it See the electric
 edition of the Journal for a color version of this figure.}]
\label{fig15}}
\end{figure}

\clearpage 

\begin{figure}
\plotone{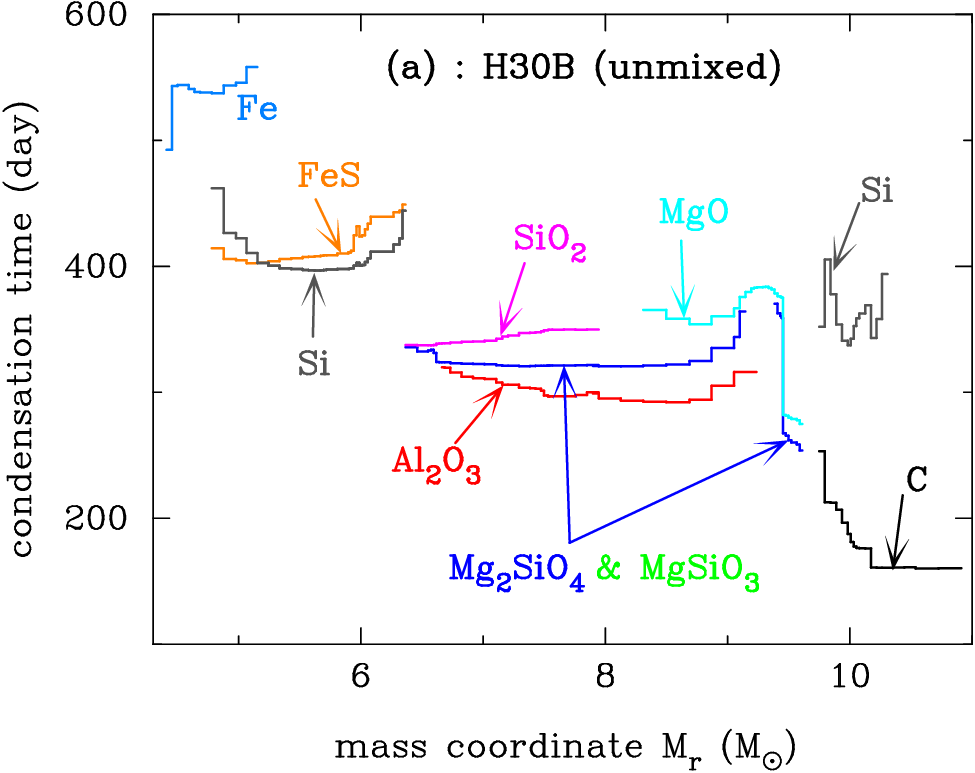}
\plotone{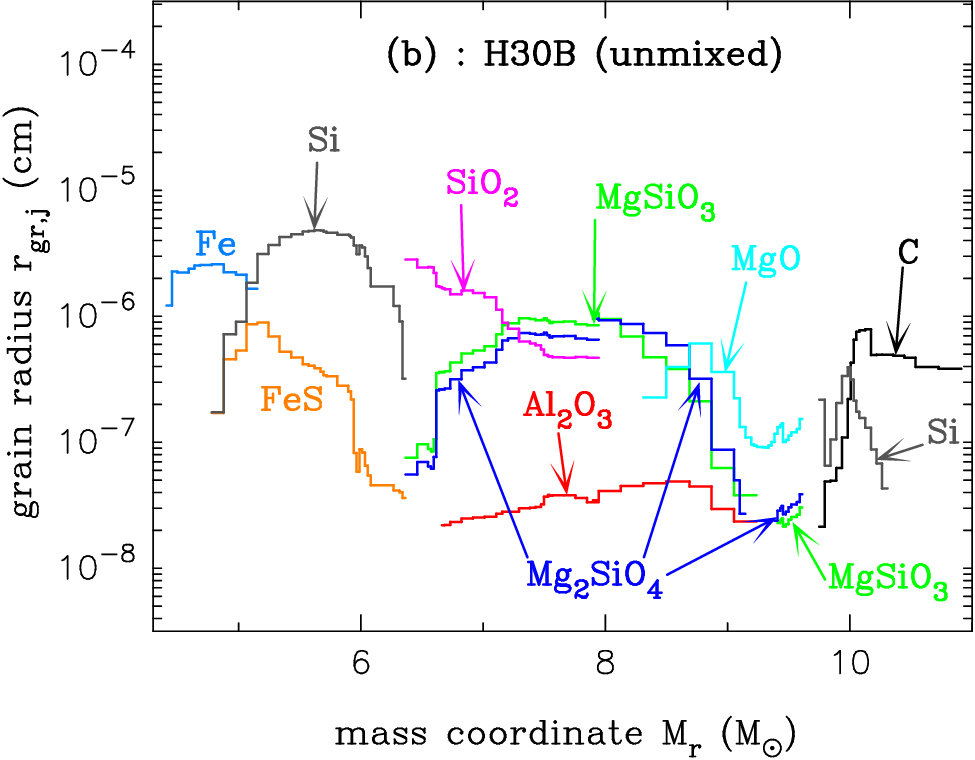}
\caption{The same as Fig 15, but for the unmixed ejecta of H30B.
[{\it See the electric
 edition of the Journal for a color version of this figure.}]
\label{fig16}}
\end{figure}

\clearpage 

\begin{figure}
\plotone{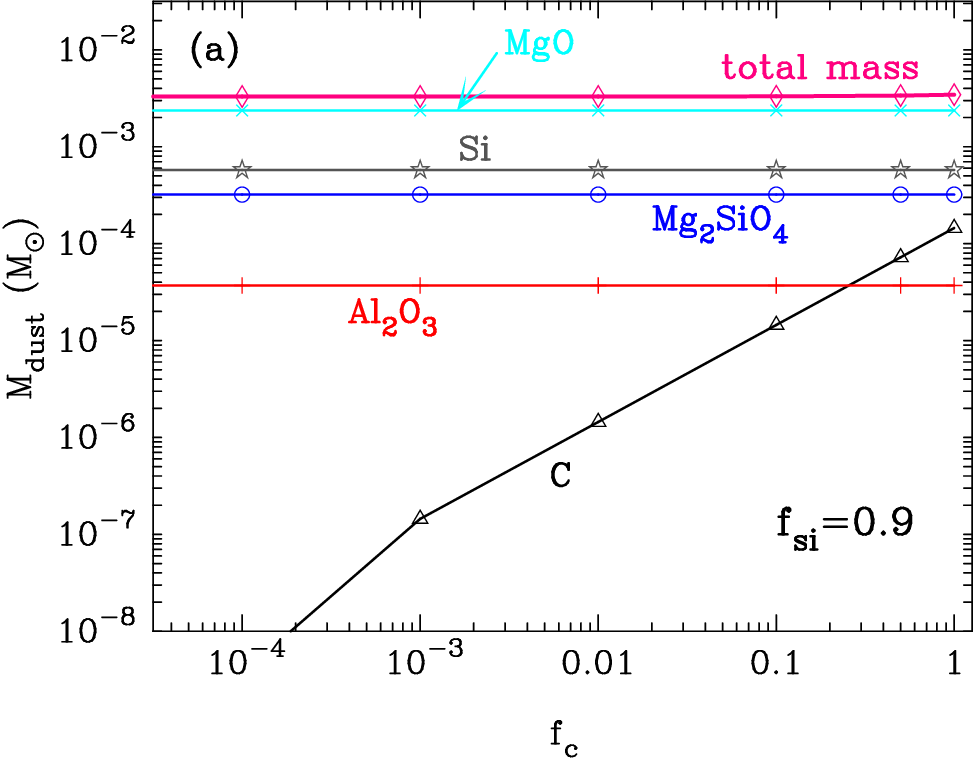}
\plotone{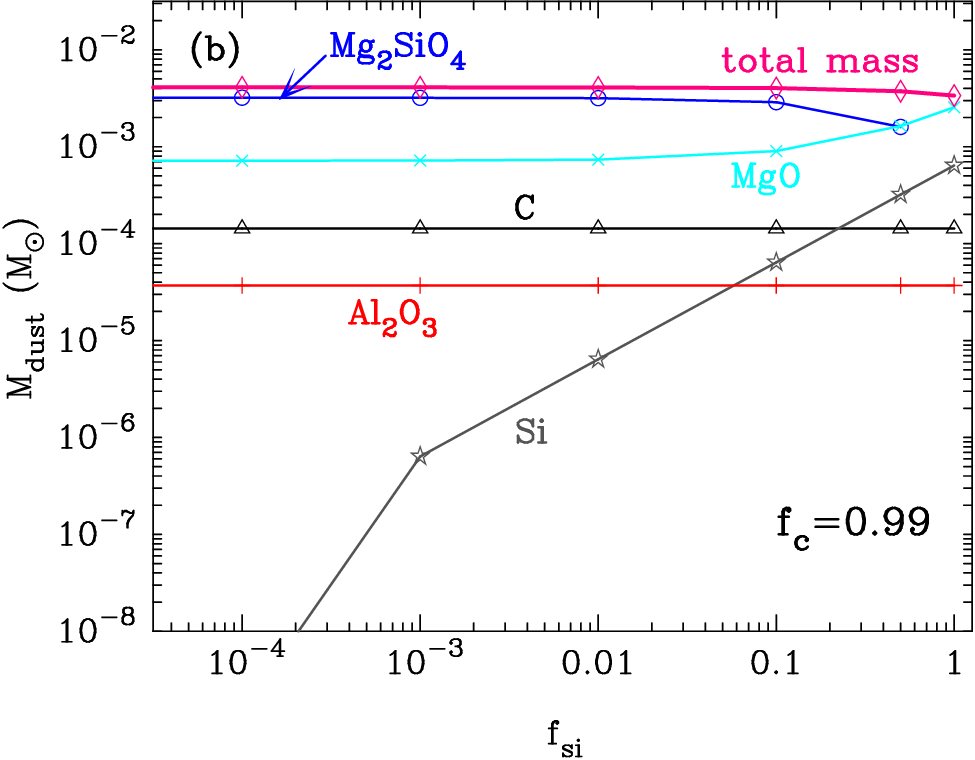}
\caption{The mass of each grain species versus the  mass
 fraction of carbon $f_{\rm C}$ and silicon $f_{\rm Si}$ not locked into
 CO and SiO molecules at $M_r =$ 3.5 $M_{\odot}$ in the O--Mg--Si layer of 
 the unmixed ejecta of 
 C20; (a) the mass of each grain species versus $f_{\rm C}$ where 
 $f_{\rm Si}$ is fixed to 0.9. (b) the mass of each grain species 
 versus $f_{\rm Si}$ with the fixed value of $f_{\rm C}=0.99$. 
}
\end{figure}

\clearpage

\begin{deluxetable}{lccccc}
\tablecaption{Models of supernovae used in the calculation. \label{tbl-2}}
\tablewidth{0pt}
\tablehead{
\colhead{Model} & \colhead{Progenitor} & \colhead{Explosion} & \colhead{Mass cut} &
\colhead{He--core} & \colhead{$M(^{56}\rm{Ni})$} \\
 & \colhead{mass} & \colhead{energy} & & \colhead{mass} &
}
\startdata
 & $M_{\odot}$ & $E_{51} (10^{51}$erg) & $M_{\odot}/M_r$ & $M_{\odot}$
 & $M_{\odot}$ \\
\tableline
C13 ..... & 13 & 1 & 1.50 & 2.05 & 0.07 \\
C20 ..... & 20 & 1 & 2.45 & 5.79 & 0.07 \\
C25 ..... & 25 & 1 & 2.54 & 8.01 & 0.07 \\
C30 ..... & 30 & 1 & 2.54 & 10.7 & 0.07 \\
P170 .... & 170 & 20 & 0 & 82.4 & 3.56 \\
P200 .... & 200 & 28 & 0 & 117.3 & 7.25 \\
\tableline
 & & & & & \\
\tableline
H25A ..... & 25 & 10 & 3.29 & 8.01 & 0.07 \\
H25B ..... & 25 & 10 & 2.47 & 8.01 & 0.7 \\
H30A ..... & 30 & 30 & 5.31 & 10.9 & 0.07 \\
H30B ..... & 30 & 30 & 4.33 & 10.9 & 0.7 \\
\enddata

\tablecomments{The labels C, P and H of the model represent ordinary 
 CCSNe, PISNe and HNe, respectively,  and the numerical value denotes the 
 mass of progenitor in units of solar mass.}

\end{deluxetable}

\clearpage

\begin{deluxetable}{lllcccc}
\tabletypesize{\scriptsize}
\tablecaption{Grain species considered in the calculation. \label{tbl-2}}
\tablewidth{0pt}
\tablehead{
\colhead{Grains} & \colhead{Key species} & \colhead{Chemical reactions} &
\colhead{$A/10^4 (\rm{K})$} & \colhead{$B$}  & $\sigma_j$ (erg/cm$^2$) &
\colhead{$a_{0j}(\rm{\AA})$}
}
\startdata
$\rm{Fe_{(s)}}$ & $\rm{Fe _{(g)}}$ & $\rm{Fe _{(g)}} \rightarrow
\rm{Fe _{(s)}} $ & 4.84180 & 16.5566 & 1800\tablenotemark{a} & 1.411 \\
$\rm{FeS_{(s)}}$ & $\rm{Fe_{(g)}/S_{(g)}}$ & $\rm{Fe_{(g)}+S_{(g)}} \rightarrow \rm{FeS _{(s)}}$ & 9.31326 & 30.7771 & 380\tablenotemark{b} & 1.932 \\
$\rm{Si_{(s)}}$ & $\rm{Si_{(g)}}$ & $\rm{Si_{(g)}} \rightarrow \rm{Si _{(s)}}$
& 5.36975 & 17.4349 & 800\tablenotemark{c} & 1.684 \\
$\rm{Ti_{(s)}}$ & $\rm{Ti_{(g)}}$ & $\rm{Ti_{(g)}} \rightarrow \rm{Ti _{(s)}}$
& 5.58902 & 16.6071 & 1510\tablenotemark{c} & 1.615 \\
$\rm{V_{(s)}}$ & $\rm{V_{(g)}}$ & $\rm{V_{(g)}} \rightarrow \rm{V _{(s)}}$ & 6.15394 & 17.8702 & 1697\tablenotemark{c} & 1.490 \\
$\rm{Cr_{(s)}}$ & $\rm{Cr_{(g)}}$ & $\rm{Cr_{(g)}} \rightarrow \rm{Cr _{(s)}}$ & 4.67733 & 16.7596 & 1880\tablenotemark{c} & 1.421 \\
$\rm{Co_{(s)}}$ & $\rm{Co_{(g)}}$ & $\rm{Co_{(g)}} \rightarrow \rm{Co _{(s)}}$ & 5.03880 & 16.8372 & 1936\tablenotemark{c} & 1.383 \\
$\rm{Ni_{(s)}}$ & $\rm{Ni_{(g)}}$ & $\rm{Ni_{(g)}} \rightarrow \rm{Ni _{(s)}}$ & 5.09130 & 17.1559 & 1924\tablenotemark{c} & 1.377 \\
$\rm{Cu_{(s)}}$ & $\rm{Cu_{(g)}}$ & $\rm{Cu_{(g)}} \rightarrow \rm{Cu _{(s)}}$ & 3.97955 & 14.9083 & 1300\tablenotemark{c} & 1.412 \\
$\rm{C_{(s)}}$ & $\rm{C_{(g)}}$ & $\rm{C_{(g)}} \rightarrow \rm{C _{(s)}}$ & 8.64726 & 19.0422 & 1400\tablenotemark{d} & 1.281 \\
$\rm{SiC_{(s)}}$ & $\rm{Si_{(g)}/C_{(g)}}$ & $\rm{Si_{(g)}+C_{(g)}} \rightarrow \rm{SiC _{(s)}}$ & 14.8934 & 37.3825 & 1800\tablenotemark{e} & 1.702 \\
$\rm{TiC_{(s)}}$ & $\rm{Ti_{(g)}/C_{(g)}}$ & $\rm{Ti_{(g)}+C_{(g)}} \rightarrow \rm{TiC _{(s)}}$ & 16.4696 & 37.2301 & 1242\tablenotemark{f} & 1.689 \\
$\rm{Al_2O_{3(s)}}$ & $\rm{Al_{(g)}}$ & $\rm{2Al_{(g)}+3O_{(g)}} \rightarrow \rm{Al_2O_{3(s)}}$ & 18.4788 & 45.3543 & 690\tablenotemark{g} & 1.718 \\
$\rm{MgSiO_{3(s)}}$ & $\rm{Mg_{(g)}/SiO_{(g)}}$ &
 $\rm{Mg_{(g)}+SiO_{(g)}+2O_{(g)}} \rightarrow \rm{MgSiO_{3(s)}}$ & 25.0129 & 72.0015 & 400\tablenotemark{h} & 2.319 \\
$\rm{Mg_2SiO_{4(s)}}$ & $\rm{Mg_{(g)}}$ &
 $\rm{2Mg_{(g)}+SiO_{(g)}+3O_{(g)}} \rightarrow \rm{Mg_2SiO_{4(s)}}$ &
 18.6200 & 52.4336 & 436\tablenotemark{h} & 2.055 \\
& $\rm{SiO_{(g)}}$ & & 37.2400 & 104.872 &  & 2.589 \\
$\rm{SiO_{2(s)}}$ & $\rm{SiO_{(g)}}$ & $\rm{SiO_{(g)}+O_{(g)}} \rightarrow \rm{SiO_{2(s)}}$ & 12.6028 & 38.1507 & 605\tablenotemark{g} & 2.080 \\
$\rm{MgO_{(s)}}$ & $\rm{Mg_{(g)}}$ & $\rm{Mg_{(g)}+O_{(g)}} \rightarrow \rm{MgO_{(s)}}$ & 11.9237 & 33.1593 & 1100\tablenotemark{g} & 1.646 \\
$\rm{Fe_3O_{4(s)}}$ & $\rm{Fe_{(g)}}$ & $\rm{3Fe_{(g)}+4O_{(g)}} \rightarrow \rm{Fe_3O_{4(s)}}$ & 13.2889 & 39.1687 & 400\tablenotemark{g} & 1.805 \\
$\rm{FeO_{(s)}}$ & $\rm{Fe_{(g)}}$ & $\rm{Fe_{(g)}+O_{(g)}} \rightarrow \rm{FeO_{(s)}}$ & 11.1290 & 31.9850 & 580\tablenotemark{g} & 1.682 \\

\enddata

\tablenotetext{a}{Elliott, Gleiser, \& Ramakrishna (1963)}
\tablenotetext{b}{Kozasa \& Hasegawa (1988)}
\tablenotetext{c}{Elliott \& Gleiser (1960)}
\tablenotetext{d}{Tabak et al. (1975)}
\tablenotetext{e}{Kozasa et al. (1996)}
\tablenotetext{f}{Rhee (1970)}
\tablenotetext{g}{Overbury, Bertrand, \& Somorjai (1975)}
\tablenotetext{h}{Boni \& Derge (1956)}

\tablecomments{The key species is defined as the gaseous species of the 
 least collisional frequency among the reactants. 
 The Gibbs free energy $\Delta G^0_j$ for formation of the condensate 
 from the reactants per the key species is approximated by a formula
 $\Delta G^0_j/kT=-A/T+B$, and the numerical values A and B are
 evaluated by least--squares fitting of the thermodynamics data (Chase
 et al. 1985) in the 
 range of temperatures of interest. $\sigma_j$ is the surface
 energy of the condensate, and $a_{0j}$ is the hypothetical radius of the
 condensate per the key species whose values are calculated from the
 molar volumes tabulated by Robie \& Waldbaum (1968).}

\end{deluxetable}

\end{document}